\documentclass[%
 reprint,
superscriptaddress,
 amsmath,amssymb,
 aps,
prb,
floatfix,
]{revtex4-2}

\usepackage{graphicx}
\usepackage{dcolumn}
\usepackage{bm}
\usepackage{physics}
\usepackage{xcolor}
\usepackage{gensymb}
\usepackage[caption=false]{subfig}

\begin{document}


\title{Low-energy Properties of Electrons and Holes in CuFeS$_2$}

\author{Bj\o rnulf Brekke}
\affiliation{Center for Quantum Spintronics, Department of Physics, NTNU - Norwegian University of Science and Technology, NO-7491 Trondheim, Norway}
\author{Roman Malyshev}
\affiliation{Department of Electronic Systems, NTNU - Norwegian University of Science and Technology, NO-7491 Trondheim, Norway}
\author{Ingeborg-Helene Svenum}
\affiliation{Department of Chemical Engineering, NTNU - Norwegian University of Science and Technology, NO-7491 Trondheim, Norway}
\affiliation{SINTEF Industry, NO-7465 Trondheim, Norway}
\author{Sverre M. Selbach}
\affiliation{Department of Materials Science and Engineering, NTNU - Norwegian University of Science and Technology, NO-7491 Trondheim, Norway}
\author{Thomas Tybell}
\affiliation{Department of Electronic Systems, NTNU - Norwegian University of Science and Technology, NO-7491 Trondheim, Norway}
\author{Christoph Br\"{u}ne}
\affiliation{Center for Quantum Spintronics, Department of Physics, NTNU - Norwegian University of Science and Technology, NO-7491 Trondheim, Norway}
\author{Arne Brataas}
\affiliation{Center for Quantum Spintronics, Department of Physics, NTNU - Norwegian University of Science and Technology, NO-7491 Trondheim, Norway}

\date{\today}

\begin{abstract}
The antiferromagnetic semiconductor CuFeS$_2$ belongs to a magnetic symmetry class that is of interest for spintronics applications. In addition, its crystal lattice is compatible with Si, making it possible to integrate it with non-magnetic semiconducting structures. Therefore, we investigate this material by finding the effective $\bm{k}\cdot\bm{p}$ Hamiltonian for the electron- and hole bands. We base this description on \textit{ab initio} calculations and classify the electronic bands by their symmetry. As a result, we find that CuFeS$_2$ exhibits spin-polarized bands and an anomalous Hall effect. Finally, we suggest using cyclotron resonance to verify our proposed effective mass tensors at the conduction band minimum and valence band maximum.
\end{abstract}

\maketitle


\section{\label{sec:introduction}Introduction}

Antiferromagnets currently attract considerable interest because of their intriguing ultrafast spin dynamics that couple to electric currents \cite{Jungwirth:NNano2016,Baltz:RMP2018,Gomonay:NPHYS2018}. The interplay between spin excitations and the transport of spin, heat, and charge can reveal novel phenomena. Recent works have demonstrated the central capabilities of antiferromagnets for use in spintronics devices. Electrical currents can switch the staggered field in antiferromagnets \cite{Wadley:SCIENCE2016,Bodnar:NatC2018,Wadley:NNAN2018}. Spins can propagate longer than micrometers in antiferromagnetic insulators \cite{Lebrun:NATURE2018}. Dynamical spins in antiferromagnets can act as spin batteries, as revealed via the inverse spin Hall effect \cite{Tserkovnyak:PRL2002,Brataas:PRB2002,Tserkovnyak:RMP2005,Cheng:PRL2014,Li:Nature2020,Vaidya:SCIENCE2020}. These features, the high-frequency capacity, and the robustness against external magnetic fields can enable new ways to realize high-speed electronics. 

Magnetic semiconductors are of interest for use in spintronics devices because of the tunable charge carrier density, integration with other semiconductors like Si and GaAs, and the possibility of creating low-dimensional structures like quantum wells, quantum wires, and quantum dots. Decades ago, dilute ferromagnetic semiconductors got attention because of the prospect of combined control of the electron spin and the charge carriers \cite{Ohno:NATURE2000,Jungwirth:RMP2006,Tanaka:JJAP2020}. Materials like GaMnAs exhibit a reasonably high Curie temperature of around 200 K, yet it is significantly below room temperature. (In,Fe)As is an electron-induced ferromagnetic semiconductor with a Curie temperature above 300 K \cite{Tanaka:JJAP2020}.

On the other hand, antiferromagnetic semiconductors are underexplored for use in spintronics devices. This class of materials combines ultra-fast spin dynamics and tunable electron- and hole properties that potentially can enable new features and reveal interesting phenomena.
CuFeS$_2$ is a good candidate because of its high Néel temperature of 823 K \cite{teranishi1961}. Additionally, its magnetic crystal structure belongs to an intriguing symmetry class. This is the type I Shubnikov class, and it generally allows for the anomalous Hall effect and spin-polarized electron bands \cite{vsmejkal2020crystal,yuan2020giant}. These properties are typical for ferromagnets. In this sense, CuFeS$_2$ may exhibit favorable properties of both ferro- and antiferromagnets. To make use of its semiconducting properties, CuFeS$_2$ can also be integrated with Si due to their compatible lattice structures \cite{kittel1996introduction}.

CuFeS$_2$ has already seen interest in different areas, mostly due to its thermoelectric properties \cite{Lazewski:PRB2004,Park:JAP2019,Hirokazu:JJAP2019,Hirokazu:APL2017,Meng:APA2015,Conejeros:IC2015, khaledialidusti2019temperature}. However, to understand its semiconductor properties, knowledge and models of the dispersions of electrons and holes at low doping levels are essential. Previous works compute, with \textit{ab initio} techniques, the electronic band structure \cite{hamajima1981self,edelbro2003full, de2010disulphide}. However, future exploration of this antiferromagnetic semiconductor requires a systematic study of the low-energy electron and hole properties while taking spin-orbit coupling into account. The purpose of the present paper is to fill this knowledge gap.

The paper is organized as follows. First, we provide a brief overview of the magnetic space group (MSG) symmetries of CuFeS$_2$. We use the MSG symmetries to give a phenomenological description of the conductivity tensor based on Neumann's principle. The following section presents \textit{ab initio} calculations of the electronic bands and the symmetry characterization of the principal bands relevant for electron- and hole transport. We then derive effective models for the valence band maximum and conduction band minimum based on this symmetry classification. Lastly, we discuss how electron cyclotron resonance can be used to verify the suggested electron- and hole band extrema and dispersions.

\section{\label{sec:symmetry}Crystal symmetry}

\begin{figure}[ht!]
\includegraphics[width=1\columnwidth]{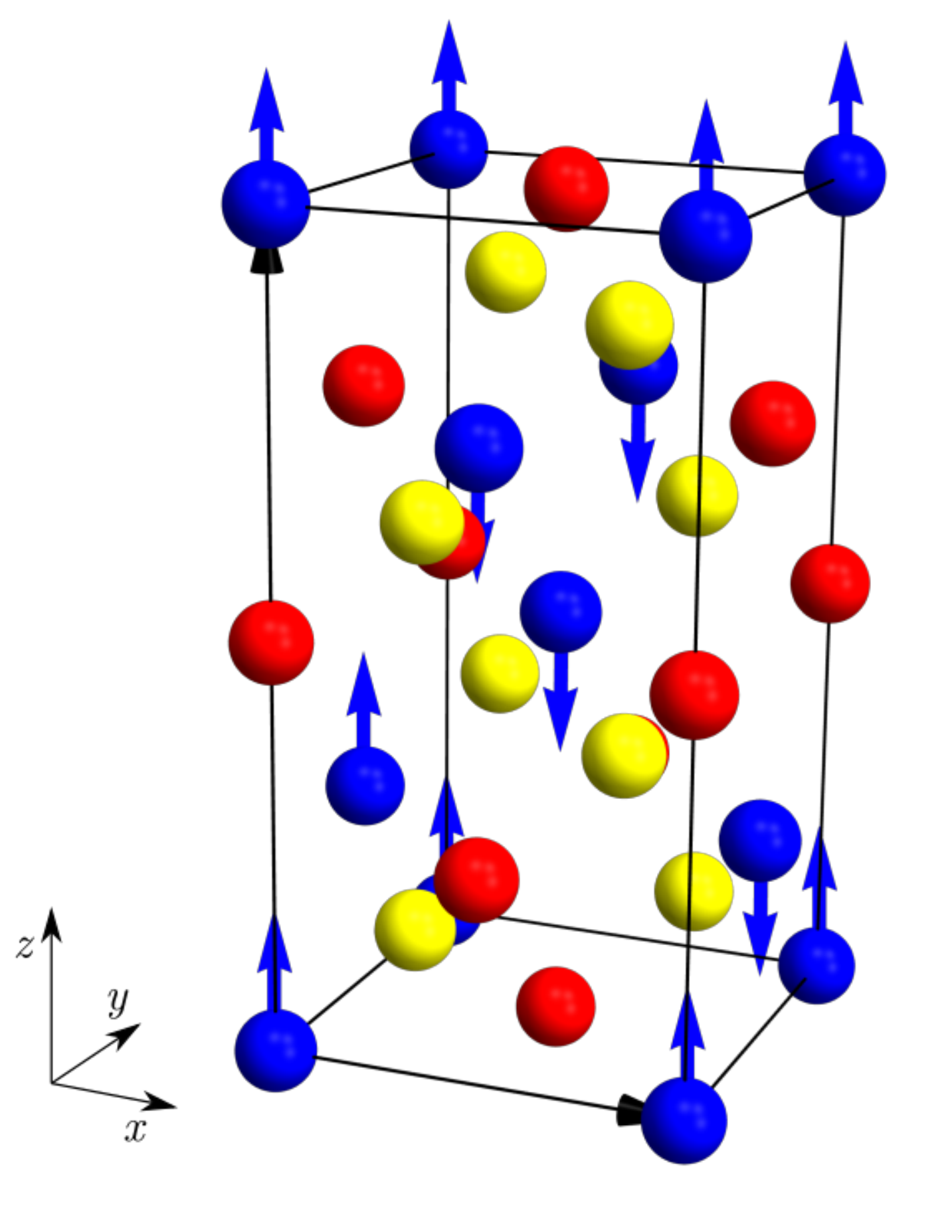}
\caption{\label{fig:wide}The conventional unit cell of CuFeS$_2$. The Fe-atoms are blue and host magnetic moments represented by arrows, the Cu-atoms are red, and the S-atoms are yellow.}
\end{figure}

Chalcopyrite, CuFeS$_2$ has a tetragonal crystal structure with space group $I\bar{4}2d$, \#122 \cite{pauling1932crystal}. To account for the collinear antiferromagnetic ground state, we consider its magnetic space group \cite{donnay1958symmetry}. This configuration is shown in Figure \ref{fig:wide}. The magnetic space group is a type I Shubnikov group, also called a Fedorov group. Such groups lack anti-unitary symmetry operations such as time-reversal symmetry. In other words, the crystal space group and the magnetic space group are isomorphic. Consequently, the chemical and the magnetic unit cell coincide. The MSG consists of the eight symmetries
\begin{subequations}
\label{eq:whole}
\begin{align}
    (E\vert0), && (C_{2z}\vert0), && (S_{4z}^+\vert 0), && (S_{4z}^-\vert 0),
    \label{eq:sym1}
\end{align}
\begin{align}
    (C_{2x}\vert \bm{\tau}), && (C_{2y}\vert \bm{\tau}), && (\sigma_{xy}\vert \bm{\tau}), && (\sigma_{\bar{x}y}\vert \bm{\tau}) , 
    \label{eq:sym2}
\end{align}
\end{subequations}
\noindent where the origin of the unit cell coincides with a Fe ion. Four of the symmetries are non-symmorphic symmetry operations. They are composite symmetries consisting of a point group operation and a translation $\bm{\tau}=(0, a/2, c/4)$ in terms of the tetragonal lattice constants $a$ and $c$. The crystal is non-centrosymmetric such that the space group allows for antiferromagnetic skyrmions \cite{Bogdanov:ZETF1989}. The non-symmorphic symmetries relate the spins of the two Fe atoms and render chalcopyrite a fully compensated antiferromagnet (AFM). The fact that the oppositely aligned spins are related by unitary non-symmorphic symmetries allows for interesting features for spintronic applications. According to the classification given by Yuan \textit{et al.} \cite{yuan2020giant}, the MSG allows for AFM-induced spin-polarized electron bands. We investigate this phenomenon further in section \ref{sec:Ab-initio calculations}. B. along with \textit{ab initio} calculations.

\section{\label{Conductivity}Electron conductivity}

The conductivity $\sigma$ captures the electron transport properties of the crystal. We consider the conductivity to first order in the Néel vector 
\begin{align}
    J_i = \sigma_{ij}E_j + \sigma_{ijk}E_j n_k.
\end{align}
Here, $i$, $j$ and $k$ refer to Cartesian directions of the current density $J_i$, the electric field $E_j$ and the Néel vector $n_k$. The first term on the right-hand side includes the second-rank conductivity tensor $\sigma_{ij}$. The second term is proportional to $n_k$ and yields a third-rank conductivity tensor $\sigma_{ijk}$. Notably, any anomalous Hall conductivity vanishes for conventional collinear AFMs. That is, AFMs with either a composite time-reversal- and inversion symmetry or time-reversal- and translation symmetry \cite{vsmejkal2020crystal}. CuFeS$_2$ has neither of these composite symmetries, which opens the possibility for a finite anomalous Hall effect.

The MSG symmetries allow for a phenomenological description of the conductivity tensors using Neumann's principle \cite{birss1966symmetry}. We require $\sigma_{ij}$ and $\sigma_{ijk}$ to be invariant under all symmetries
\begin{subequations}
\begin{align}
    \sigma_{ij} = \mathcal{R}_{ii'}\mathcal{R}_{jj'}\sigma_{i'j'},
\end{align}
\begin{align}
    \sigma_{ijk} = (-1)^{l+m}\mathcal{R}_{ii'}\mathcal{R}_{jj'}\mathcal{R}_{kk'}\sigma_{i'j'k'}.
\end{align}
\end{subequations}
The Néel vector $n_k$ has special transformation properties. It is a pseudovector which means it is invariant under orientation reversal. Also, it changes sign under sublattice exchange caused by the non-symmorphic transformations. To account for this, we introduce the boolean variables $l,m \in \{0,1\}$, which are non-zero for orientation-reversing- and non-symmorphic symmetries, respectively.

As a result, the conductivity relations to the first order in $\bm{n}$ are
\begin{subequations}
\begin{align}
    J_x &= \sigma_t E_x + \sigma_t' E_x n_z + \sigma_{A}'E_z n_x, \\
    J_y &= \sigma_t E_y + \sigma_t' E_y n_z + \sigma_{A}'E_z n_y, \\
    J_z &= \sigma_l E_z + \sigma_l' E_z n_z + \sigma_{B}'(E_x n_x + E_y n_y).
\end{align}
\label{relations}
\end{subequations}
The coefficients $\sigma_t, \sigma_l, \sigma_l'$, $\sigma_t'$, $\sigma_A'$ and $\sigma_B'$ can be found empirically or from microscopic calculations.
The relations in Eq. \eqref{relations} allow for direct measurement of both sign and direction of the Néel vector. In the following, we consider the electron structure properties using \textit{ab initio} calculations.

\section{\label{sec:Ab-initio calculations}Ab-initio calculations}
 
\subsection{\label{subsec:Methods}Methods}

Density functional theory (DFT) calculations were performed with the projected augmented-wave method \cite{Blochl:PRB1994, KresseJoubert:PRB1999}. Accordingly, we used the Vienna \emph{Ab Initio} Simulation Package (VASP) in the DFT+U methodology, including spin-orbit coupling (SOC), to capture the SOC-induced band splitting at high-symmetry points. To represent the bulk crystal, we used a periodic model of a single conventional unit cell that contains 16 atoms. Furthermore, we sampled the Brillouin zone using a $\Gamma$-centered mesh with at least $4\times4\times2$ $k$-points. The mesh was generated with the Monkhorst-Pack scheme \cite{MonkhorstPack:PRB1976}. For the density of states (DOS) calculations, we changed the $k$-mesh to include $12\times12\times6$ $k$-points. The electron band structure was sampled at 100 $k$-points on the interval between each pair of high-symmetry points on the $\bm{k}$-path. We set an energy cut-off at 700 eV.

To relax the structure and minimize the total energy, we applied the iterative conjugate-gradient method. We relaxed the atomic positions until the residual forces acting on the atoms were smaller than 10\textsuperscript{-5} eV/Å and the energy difference in the final convergence step was smaller than 10\textsuperscript{-8} eV per unit cell.

The Fe magnetic moments converged to consistent values for initial inputs on the interval 1 to 6 Bohr magnetons $\mu_B$. The spins are oriented antiferromagnetically, as illustrated
in Figure \ref{fig:wide}. We considered other collinear AFM orderings, as well as a ferromagnetically ordered structure. In consistence with the results of Ref. \cite{Conejeros:IC2015},  these orderings produced a higher energy state than the chosen AFM structure.

\begin{figure}[ht!]
\includegraphics[width=1\columnwidth]{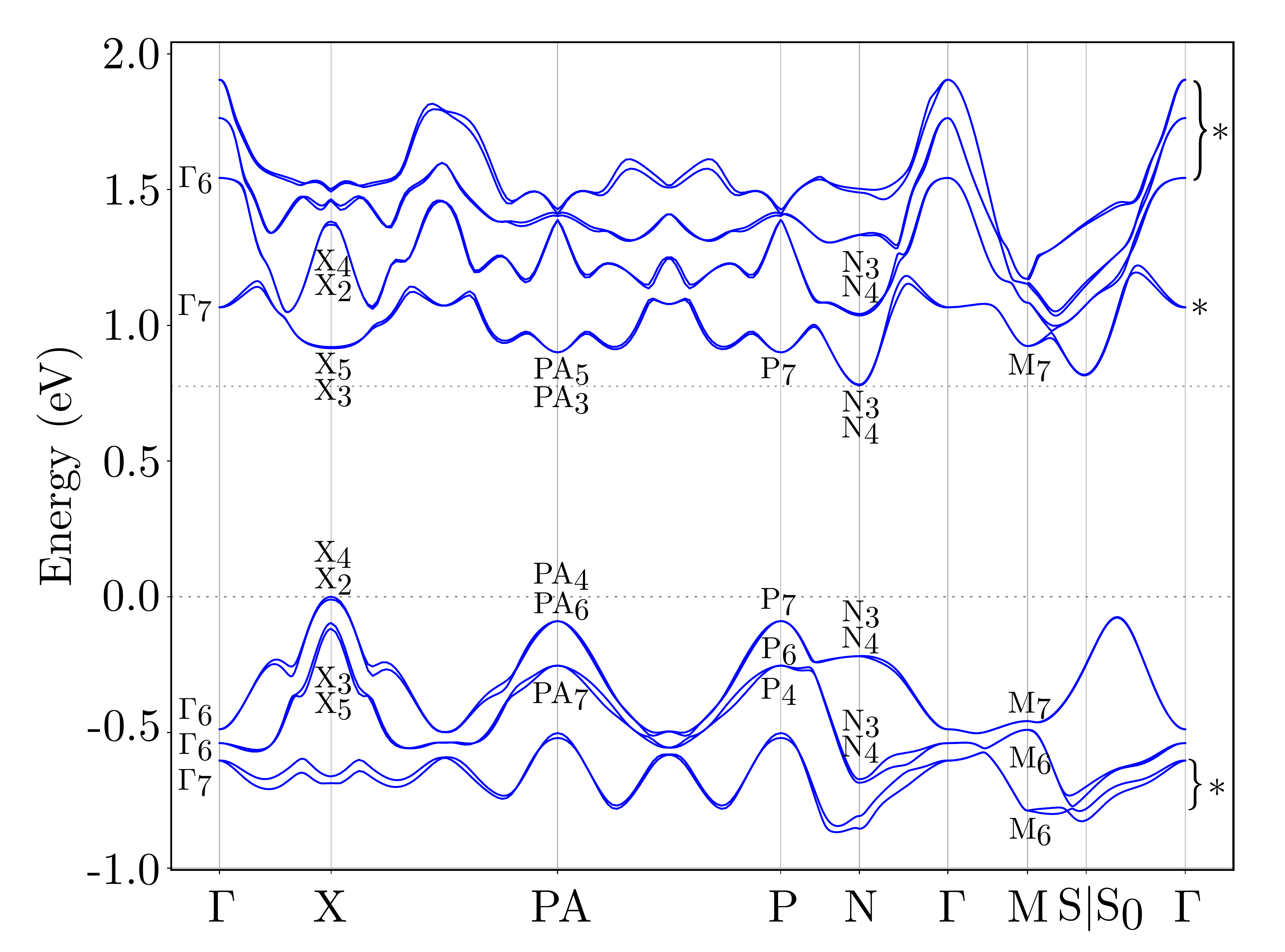}
\caption{\label{fig:bandStructure}The electronic band structure of CuFeS$_2$ with spin-orbit coupling. The bands are classified by irreducible representations at the high-symmetry points according to the notation in Ref. \cite{elcoro2017double}. Isolated bands marked by an asterisk have a non-trivial topological index.}
\end{figure}

\begin{figure}[ht!]
\centering
\begin{subfloat}[\label{subfig:ados}]{\includegraphics[width=1\columnwidth]{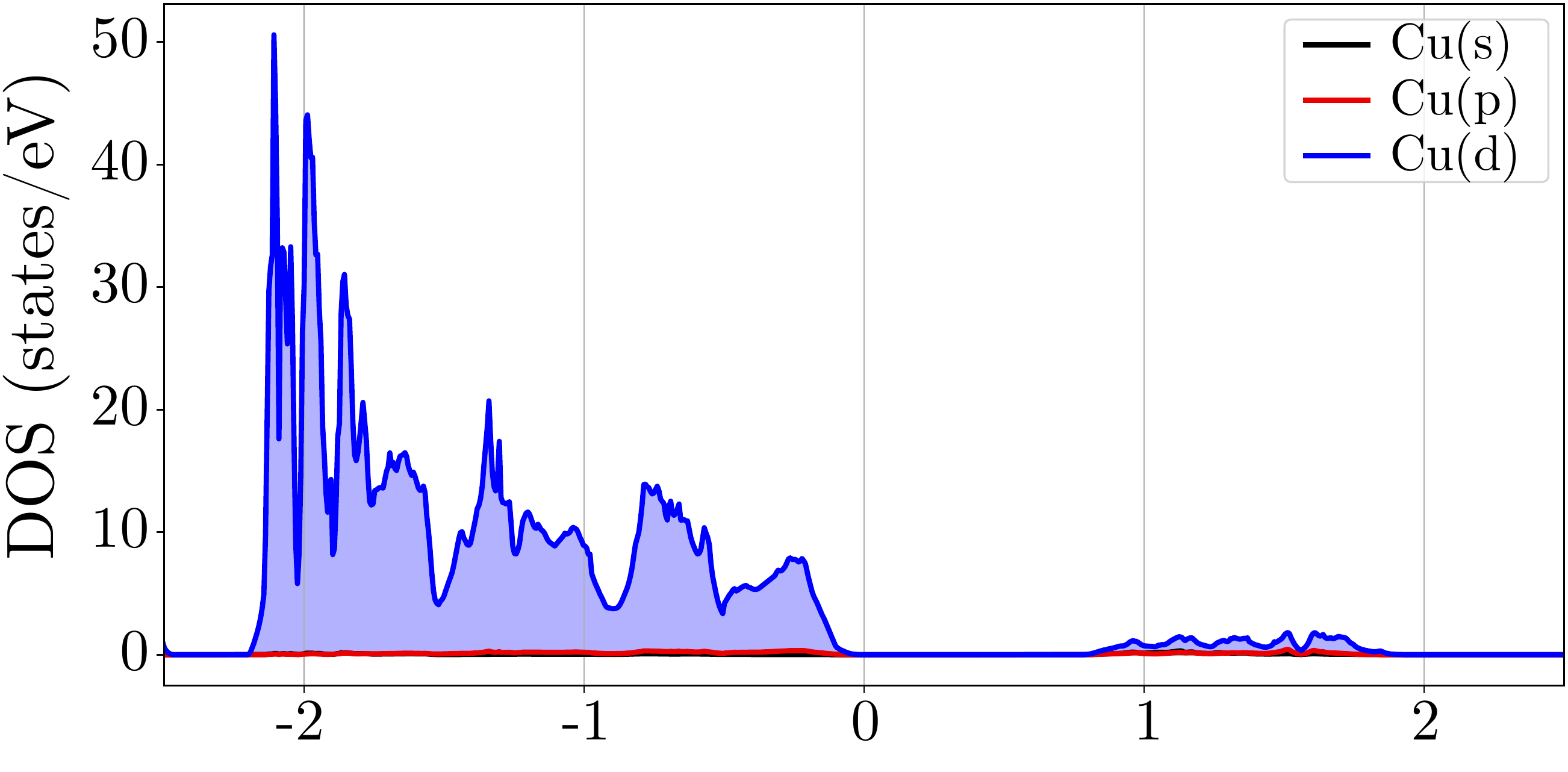}}
\end{subfloat}
\begin{subfloat}[\label{subfig:bdos}]{\includegraphics[width=1\columnwidth]{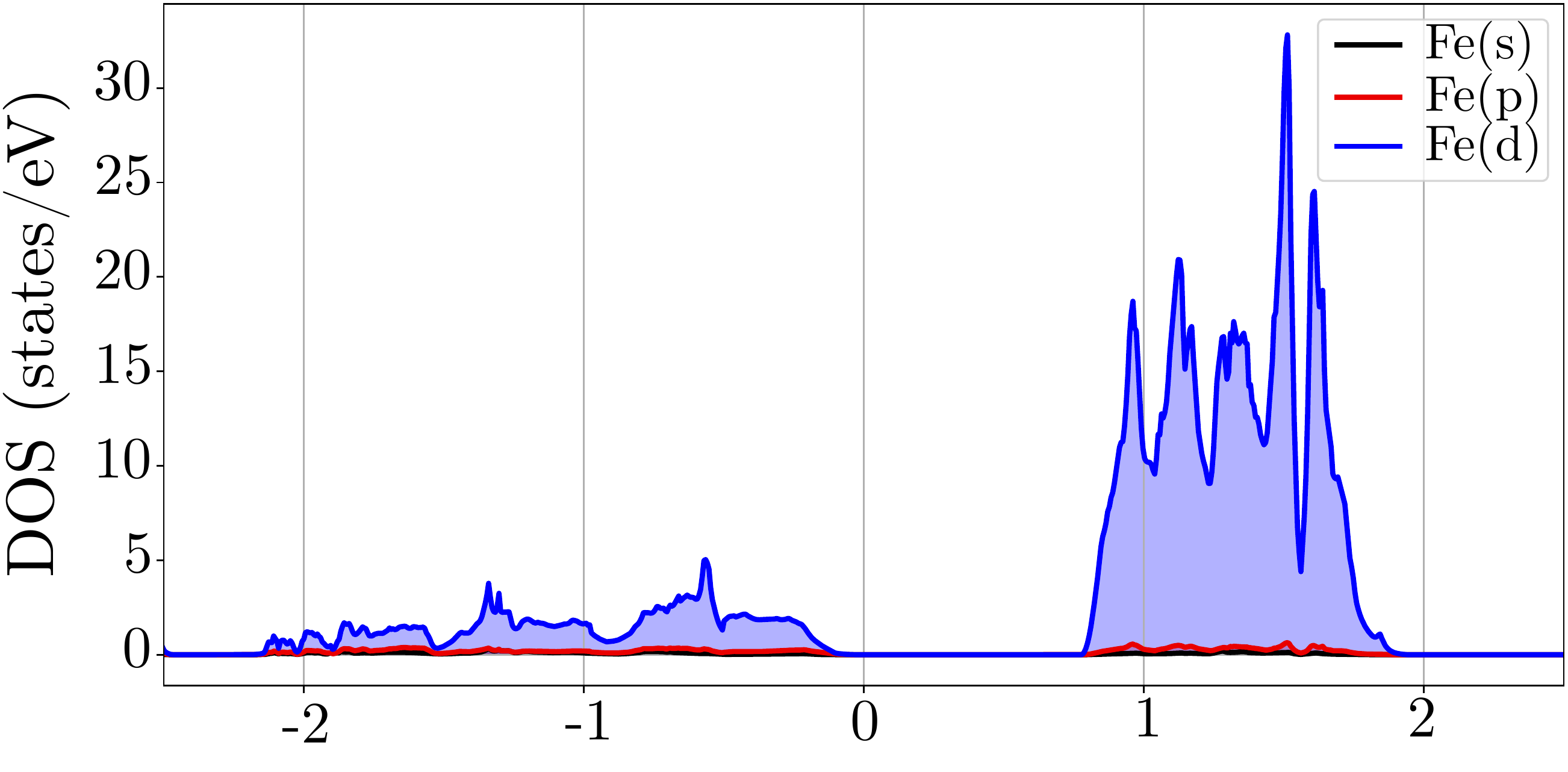}}
\end{subfloat}
\begin{subfloat}[\label{subfig:cdos}]{\includegraphics[width=1\columnwidth]{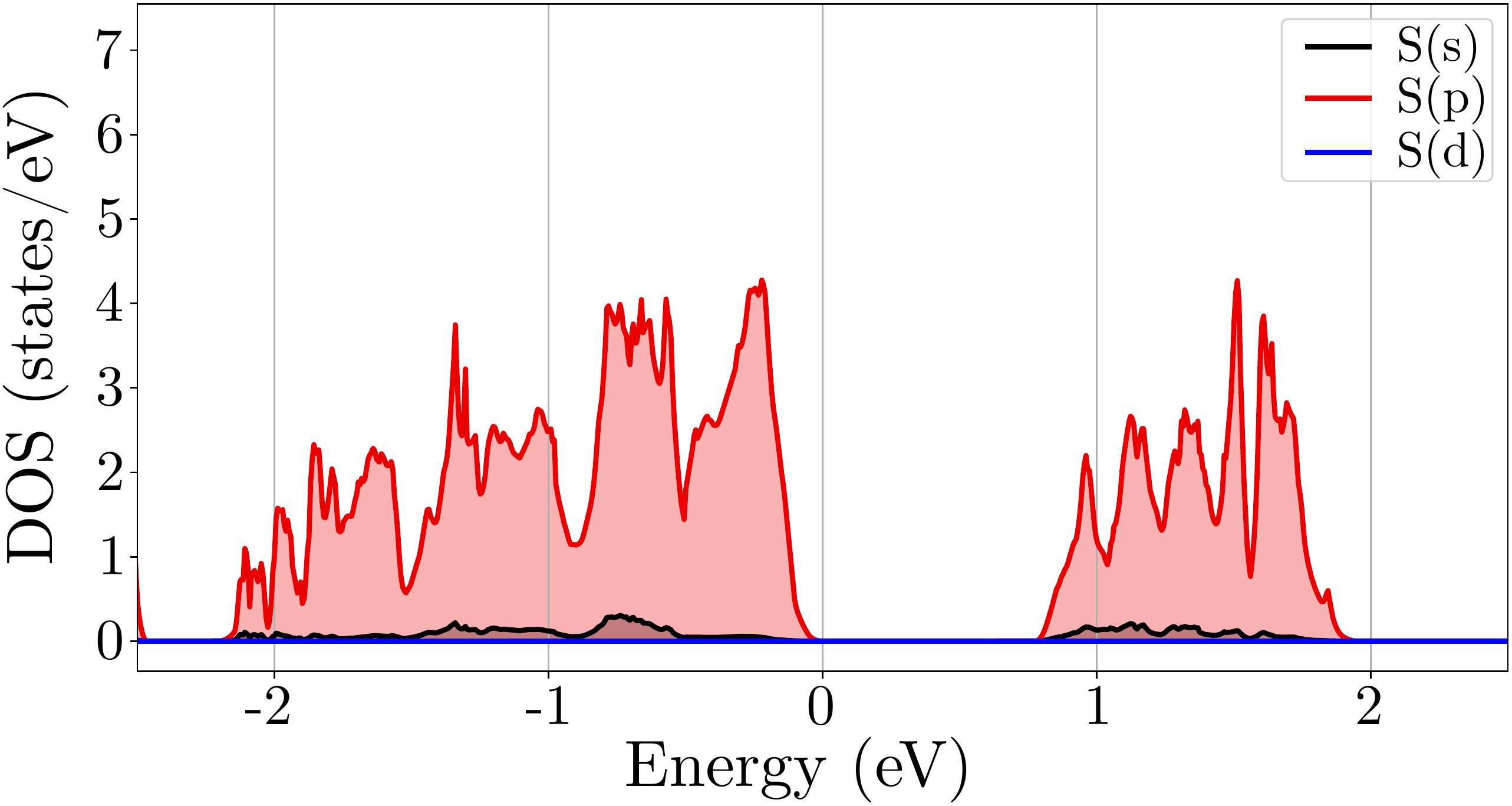}}
\end{subfloat}
\caption{\label{fig:DOS}The atom-decomposed density of state from \textit{ab initio} calculations of CuFeS$_2$. The Cu $3d$- and S $3p$ orbitals dominate the valence band, whereas the Fe $3d$- and S $3p$ orbitals dominate the conduction band. The scaling on the $y$-axes differs for better visibility.} 
\end{figure}

The DFT+U scheme of Dudarev \textit{et al.} \cite{Dudarev:PRB1998} was used, with a varying $U_{\text{eff}}$ applied to Fe $3d$ and a $U_{\text{eff}}=0.1$ eV applied to Cu $3d$.
We used the Perdew-Burke-Ernzerhof functional for solids (PBEsol) \cite{PBEsol} for the relaxation, electronic structure, density of states, and band structure calculations. The states $3s^2 3p^4$, $3p^6 4s^1 3d^{10}$ and $3p^6 4s^1 3d^7$ were treated as valence electrons for the atomic types S, Cu and Fe, respectively. To determine a fitting value of $U_{\text{eff}}$ in DFT+U, we performed a full relaxation with the HSE06 hybrid functional \cite{HSE06}. Table \ref{tab:functionalscomparison} presents the results from the PBEsol relaxations with various $U_{\text{eff}}$ values applied to Fe 3$d$ orbitals, as well as those determined with the HSE06 functional. We fit the $U_{\text{eff}}$ value to the first lattice constant, $a$ and calculated the DOS and band structure with $U_{\text{eff}}=4.7$. This value gives the lattice constants $a = 5.259$ Å and $c = 10.407$ Å.

\begin{table}[b]
\caption{\label{tab:functionalscomparison}Lattice parameters $a$, $c$, their ratio $c/a$ and magnetic moment of Fe atoms for DFT+U models using PBEsol with different $U_{\text{eff}}$ values for Fe 3$d$. The bottom row shows the same quantities calculated with the hybrid functional HSE06 for comparison.}
\begin{ruledtabular}
\begin{tabular}{cccccc}
$U_{\text{eff}}$ (eV)& $a$ (Å) & $c$ (Å) & $c/a$ & $m_s$ ($\mu_{B}$) \\
\hline
0 & 5.152 & 10.177 & 1.975 & 2.620 \\
1 & 5.196 & 10.225 & 1.968 & 3.127 \\
2 & 5.218 & 10.285 & 1.971 & 3.335 \\
3 & 5.236 & 10.337 & 1.974 & 3.487 \\
4 & 5.251 & 10.380 & 1.977 & 3.607 \\
5 & 5.263 & 10.416 & 1.979 & 3.709 \\
6 & 5.275 & 10.442 & 1.980 & 3.801 \\
7 & 5.301 & 10.420 & 1.966 & 3.883 \\
8 & 5.305 & 10.473 & 1.974 & 3.966 \\
9 & 5.288 & 10.571 & 1.999 & 4.051 \\
\hline
HSE06 & 5.259 & 10.366 & 1.971 & 3.715 \\
\end{tabular}
\end{ruledtabular}
\end{table}

\subsection{\label{subsec:Results}Results}
\begin{figure}[ht!]
\begin{subfloat}[\label{subfig:a}]{\includegraphics[width=1\columnwidth]{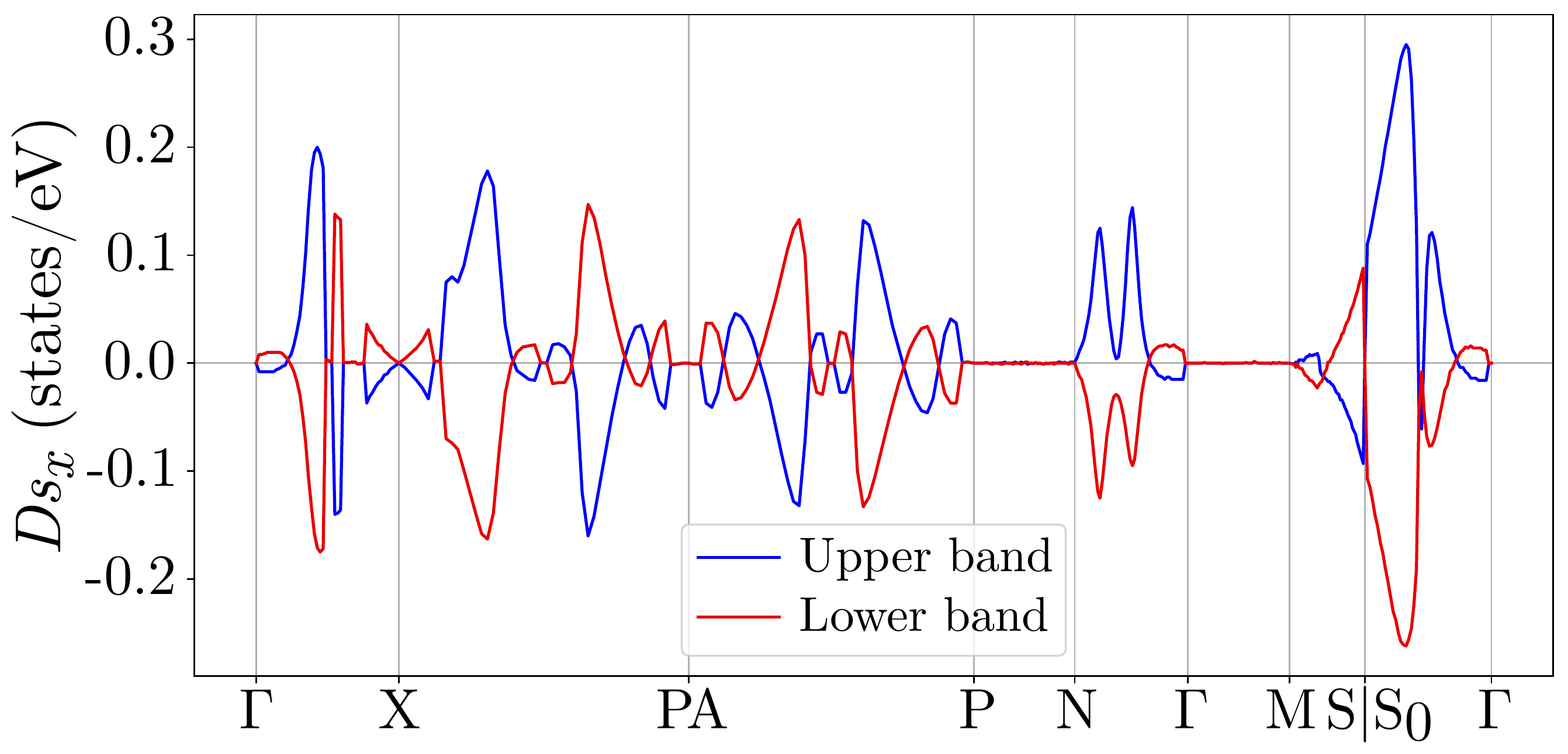}}
\end{subfloat}
\begin{subfloat}[\label{subfig:b}]{\includegraphics[width=1\columnwidth]{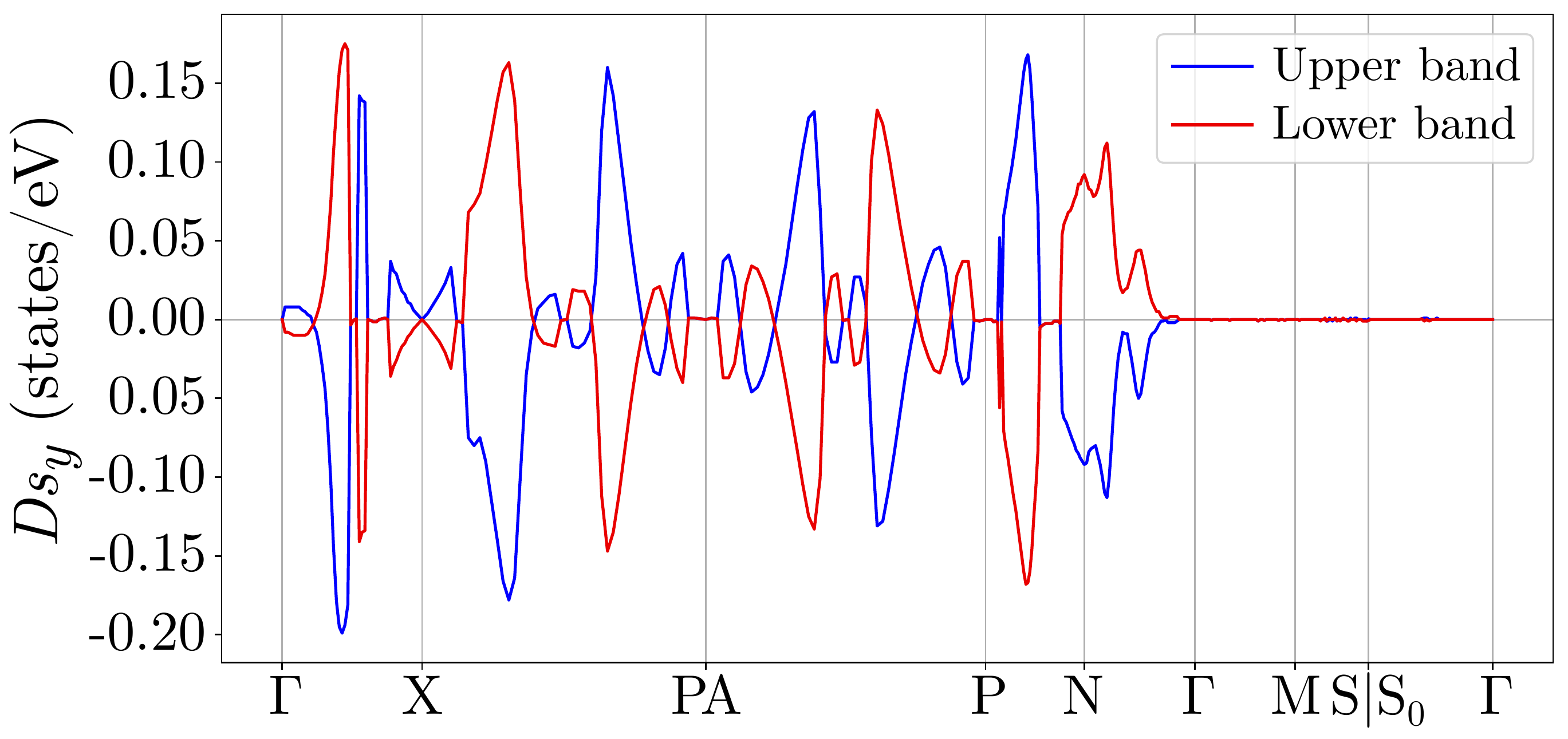}}
\end{subfloat}
\begin{subfloat}[\label{Fig:SpinZ}]{\includegraphics[width=1\columnwidth]{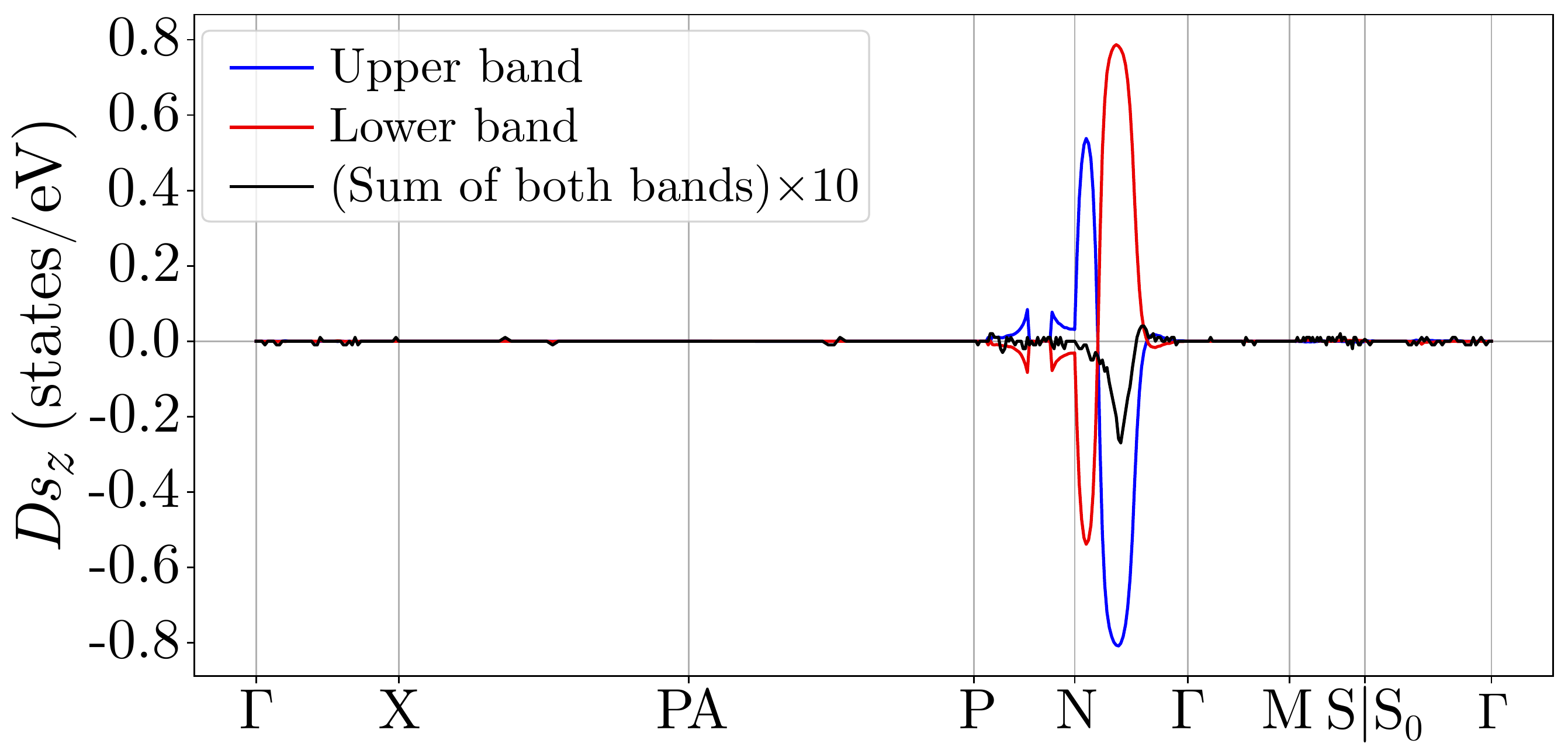}}
\end{subfloat}
\caption{\label{fig:spinPolarization}The net spin projected density of states $Ds_i$ in the (a) $i=x$-, (b) $i=y$- and (c) $i=z$ -direction of the two lowest electron conduction bands. The numerical results are consistent with the crystal symmetry, except for the partly finite spin polarization in the $z$-direction on the P-N interval.} 
\end{figure}

Figure \ref{fig:bandStructure} shows the calculated electronic band structure. The bands follow a path in $\bm{k}$-space traversing the high-symmetry points denoted in Table \ref{tab:HighSymmetryPoints}. Consistent with other \textit{ab initio} calculations \cite{Park:JAP2019}, the band structure is strongly anisotropic. The valence band maximum (VBM) is at the X-point, and the conduction band minimum (CBM) is at the N-point. The X-point is part of a two-fold valley structure, and the N-point is part of a four-fold valley structure. This renders chalcopyrite a multivalley semiconductor. The bands show an indirect band gap of 0.779 eV. Experimentally, values between 0.3-0.6 eV have been reported \cite{goodman1954, enginPowellHull2011}. The direct band gap is 0.915 eV and found at the X-point.

Figure \ref{fig:DOS} shows the atom-decomposed electron density of states. Cu- and S electrons dominate the valence band. On the other hand, the conduction bands are dominated by Fe- and S electrons. These findings suggest that conduction band electrons are more strongly coupled to the localized magnetic moments than the valence band electrons are. Hence, the spin polarization of the conduction band is of special interest.

Because the sublattice spins of CuFeS$_2$ are related by non-symmorphic symmetries instead of composite time-reversal and inversion symmetry, its magnetic symmetry class allows for spin-polarized electron bands. Figure \ref{fig:spinPolarization} shows the net spin projected density of states for the two lowest conduction bands on the considered $\bm{k}$-path based on the \textit{ab initio} calculations. At specific points, the crystal symmetries prohibit spin-polarizations in all directions. This is the case for the $\Gamma$-, X-, PA-, P-, and M-point, as well as for some high-symmetry lines. Notably, the CBM at the N-point exhibits a finite spin-polarization in the $y$-direction. The spin polarization of the related valley N$'$ has an opposite sign. There is a similar relationship between N$''$ and N$'''$ with spin polarization in the $x$-direction.

Spin polarization in the $z$-direction is more restricted by symmetry. The crystal symmetry requires it to be zero at all high-symmetry points and -lines with a non-symmorphic symmetry as part of their little group. For the chosen $\bm{k}$-path, only the N-$\Gamma$ interval lack the symmetries that prohibit a finite spin $z$ polarization. This is the only interval that is not a high-symmetry line. Although the interval is a small part of this specific $\bm{k}$-path, it represents most of the Brillouin zone in the sense that its little group contains the identity operation only. Hence, we expect a large part of the Brillouin zone to exhibit electron bands spin-polarized in the $z$- direction as well as for the $x$- and $y$-direction.

Large efforts have been made to determine the topological character of magnetic compounds. Based on the \textit{ab initio} calculations and magnetic quantum chemistry software \cite{xu2020high,elcoro2021magnetic}, we find that certain bands of CuFeS$_2$ are topologically non-trivial with respect to a $\mathbb{Z}_2$ topological invariant. These isolated bands are marked by an asterisk in Figure \ref{fig:bandStructure}. The total topological index summed over all occupied bands is zero. Hence, the \textit{ab initio} calculations and the crystal symmetries suggest that CuFeS$_2$ does not exhibit topologically protected edge states.

\begin{table}[b]
\caption{\label{tab:HighSymmetryPoints}
High-symmetry points and -lines of the conventional Brillouin zone of chalcopyrite and the little group of the wave vector. Primed symmetries are non-symmorphic and include the translation $\bm{\tau}=(0, a/2, c/4)$.}
\begin{ruledtabular}
\begin{tabular}{ccc}
$\Gamma$& $(0,0,0)$ &$EC_{2z}S_{4z}^+S_{4z}^-C_{2x}'C_{2y}'\sigma_{xy}'\sigma_{\bar{x}y}'$ \\
M& $(0,0,1)$ & $EC_{2z}S_{4z}^+S_{4z}^-C_{2x}'C_{2y}'\sigma_{xy}'\sigma_{\bar{x}y}'$ \\
P& $(1/2, 1/2, 1/2)$ & $EC_{2z}S_{4z}^+S_{4z}^-C_{2x}'C_{2y}'\sigma_{xy}'\sigma_{\bar{x}y}'$\\
PA& $(1/2, 1/2, -1/2)$ & $EC_{2z}S_{4z}^+S_{4z}^-C_{2x}'C_{2y}'\sigma_{xy}'\sigma_{\bar{x}y}'$ \\
X& $(1/2,1/2,0)$ & $EC_{2z}\sigma_{xy}'\sigma_{\bar{x}y}'$ \\
N& $(1/2,0,1/2)$ & $EC_{2y}'$
\end{tabular}
\end{ruledtabular}
\end{table}

\section{\label{sec:Effective theory}Effective mass theory}
In this section, we develop an effective theory for the band structure of the CBM and VBM. To that end, we use $\bm{k}\cdot\bm{p}$ theory \cite{voon2009kp, dresselhaus2007group}.

We start with the Hamiltonian
\begin{equation}
    \mathcal{H}=\frac{p^2}{2m}+V(\bm{r})+\frac{\hbar}{2m^2c^2}(\nabla V(\bm{r}) \times \bm{p})\cdot \bm{\sigma},
    \label{eq:Ham}
\end{equation}
where $\bm{p}$ is the momentum operator and $m$ is the electron mass. The crystal potential $V(\bm{r})$ gives rise to SOC. The SOC term contains the spin Pauli matrices $\bm{\sigma}$.

Following the standard approach, we let the Hamiltonian in Eq. \eqref{eq:Ham} act on Bloch waves $\psi_{n\bm{k}}(\bm{r})=e^{i\bm{\kappa}\cdot\bm{r}}\left(e^{i\bm{k}_0\cdot\bm{r}}u_{n\bm{k}_0}({\bm{r}})\right)$, where the wave function $u_{n\bm{k_0}}({\bm{r}})$ has the lattice periodicity. The crystal momentum $\bm{k}=\bm{k}_0 + \bm{\kappa}$, where $\bm{\kappa}$ is a small deviation from a high-symmetry point $\bm{k}_0$. We consider the small deviation perturbatively and write
\begin{equation}
    \mathcal{H} = \mathcal{H}_{\bm{k}_0} + \mathcal{H}_{\bm{\kappa}\cdot\bm{p}}',
    \label{pert}
\end{equation}
where 
\begin{equation}
    \mathcal{H}_{\bm{k}_0}=\frac{p^2}{m^2}+V(\bm{r}) + \frac{\hbar}{4m^2c^2}\left(\nabla V \times (\bm{p}+\bm{k}_0)\right)\cdot \bm{\sigma}
    \label{Hk0}
\end{equation}
and
\begin{equation}
    \mathcal{H}_{\bm{\kappa}\cdot\bm{p}}'=\frac{\hbar}{m}\bm{\kappa}\cdot\bm{P}.
    \label{Hkp}
\end{equation}
Here, $\bm{P}$ is a generalized momentum defined as
\begin{equation}
    \bm{P} = \left(\bm{p}+\frac{\hbar}{4mc^2}\bm{\sigma}\times\nabla V\right)
\end{equation}
and transforms as a radial vector.

\subsection{\label{subsec:Xpoint}Valence band maximum}

The valence band maximum is located at the $\mathrm{X}=2\pi(1/2a, 1/2a, 0)$ point. It is part of a two-valley structure together with $\mathrm{X}'=2\pi(-1/2a,1/2a,0)$. These points are related through symmetry transformations. Table \ref{tab:HighSymmetryPoints} shows the group of the wave vector. We characterize the electron bands by double group representations. At the X-point, there are four one-dimensional irreducible representations (irreps). These couple through the radial vector $\bm{P}$ as shown in Table \ref{Tab:MatrixProductsX}.

\begin{table}[ht]
\caption{Selection rules for the electronic states at the X-point.}
\centering
\begin{ruledtabular}
\begin{tabular}{ p{8cm} }
\vspace{-7mm}
{\begin{align*}
 \frac{\hbar}{m}\bra{u^{X_i}}(P_x, P_y, P_z)\ket{u^{X_i}} &= (0, 0 , a_{ii})  \\
 \frac{\hbar}{m}\bra{u^{X_2}}(P_x, P_y, P_z)\ket{u^{X_3}} &= (a_{23}, a_{23}, 0) \\
 \frac{\hbar}{m}\bra{u^{X_2}}(P_x, P_y, P_z)\ket{u^{X_4}} &= (0, 0, 0) \\
 \frac{\hbar}{m}\bra{u^{X_2}}(P_x, P_y, P_z)\ket{u^{X_5}} &= (a_{25}, -a_{25}, 0)  \\
 \frac{\hbar}{m}\bra{u^{X_3}}(P_x, P_y, P_z)\ket{u^{X_4}} &= (a_{34}, -a_{34}, 0) \\
 \frac{\hbar}{m}\bra{u^{X_3}}(P_x, P_y, P_z)\ket{u^{X_5}} &= (0, 0, 0) \\
 \frac{\hbar}{m}\bra{u^{X_4}}(P_x, P_y, P_z)\ket{u^{X_5}} &= (a_{45}, a_{45}, 0)
 \end{align*}}
\vspace{-4mm}
\end{tabular}
\label{Tab:symmetries}
\end{ruledtabular}
\label{Tab:MatrixProductsX}
\end{table}
Figure \ref{fig:bandStructure} shows that the valence band maximum consists of two non-degenerate bands. The upper and lower band transform as the X$_4$ and X$_2$ irreps, respectively. They do not couple through the generalized momentum $\bm{P}$ as shown in Table \ref{Tab:MatrixProductsX}. The splitting of these bands is 10.2 meV and induced by SOC.
To describe the bands, we consider a two-band model based on Eq. \eqref{pert}. In deriving the secular equation, we consider second-order contributions outside the two nearly-degenerate bands X$_2$ and X$_4$. The first-order intra-band couplings are zero at band extrema.

The resulting secular equation is
\begin{equation}
    \mathcal{H}^{\mathrm{X}}_{2\times2}\bm{\psi} = E\bm{\psi},
    \label{hamXPoint}
\end{equation}
where the two-component $\bm{\psi}$ denotes the wave function coefficients. The free parameters quantifying the inter-band coupling are defined in Appendix \ref{sec:DefinitionOfFreeParameters}.1. The effective Hamiltonian $\mathcal{H}^{\mathrm{X}}_{2\times2}$ is
\begin{widetext}
\begin{subequations}
\begin{align}
    \mathcal{H}_{11}=&E_2  + (A_2+C_2)(\kappa_x^2 + \kappa_y^2) + B_2\kappa_z^2 + 2(A_2-C_2)\kappa_x\kappa_y + \frac{\hbar^2}{2m}\kappa^2, \\ \mathcal{H}_{12} =& (M+L)(\kappa_x^2 - \kappa_y^2), \\ \mathcal{H}_{21} =& (M^*+L^*)(\kappa_x^2-\kappa_y^2), \\ \mathcal{H}_{22} =& E_4  + (A_4+C_4)(\kappa_x^2 + \kappa_y^2) + B_4\kappa_z^2  -2(A_4-C_4)\kappa_x\kappa_y + \frac{\hbar^2}{2m}\kappa^2.
\end{align}
\label{Xmodel}
\end{subequations}
\end{widetext}
The Hamiltonian $\mathcal{H}^{\mathrm{X}}_{2\times2}$ acts on a pair of wave functions transforming as the representations X$_2$ and X$_4$. These wave functions have contributions from all three types of atoms, whereas the largest contributions are from the Cu-ions. For a more intuitive understanding of these wave functions, we consider basis functions of the irreps with the relevant orbital characters. That is, we consider functions of identical transformation properties, under the little group symmetry operations, as the electron wave functions. We present the basis functions on the form
\begin{align}
    \psi = \ket{\text{A}, \bm{k}_0, \text{O}, n}\begin{pmatrix}\alpha \\ \beta\end{pmatrix},
\end{align}
where A denotes atom type, $\bm{k}_0$ denotes the high-symmetry point, O denotes the orbital part of the wave function, and $n\in \{1,2\}$ the location in the primitive unit cell such that $n=1$ and $n=2$ are related by the non-symmorphic translation. The last factor is the spinor. A similar treatment has been done for the P-point \cite{Mikhailovskii:PSSB1990}. The S-ion part of the X$_2$ state is of p-orbital character, and the Cu-ion part is mostly of d-orbital character. The p-orbital basis functions are of the form
\begin{subequations}
\begin{align}
\begin{split}
    \psi_{p_x+ip_y} =& \ket{\mathrm{S},\mathrm{X}, x+iy, 1}\begin{pmatrix}1 \\ 0\end{pmatrix} \\&+e^{i\frac{3\pi}{4}}\ket{\mathrm{S}, X, x-iy, 2}\begin{pmatrix}0 \\ 1\end{pmatrix},
\end{split}\\
\begin{split}
    \psi_{p_x-ip_y} =& \ket{\mathrm{S},\mathrm{X}, x-iy, 1}\begin{pmatrix}1 \\ 0\end{pmatrix} \\&-e^{i\frac{3\pi}{4}}\ket{\mathrm{S}, X, x+iy, 2}\begin{pmatrix}0 \\ 1\end{pmatrix},\end{split}
    \\ \psi_{p_z} =& \ket{\mathrm{S},\mathrm{X}, z, 1}\begin{pmatrix}0 \\ 1\end{pmatrix} +e^{\frac{3\pi}{4}i} \ket{\mathrm{S},\mathrm{X}, z, 2}\begin{pmatrix}1 \\ 0\end{pmatrix}.
\end{align}
\label{BasisfunctionsXS}
\end{subequations}
The dominant d-orbitals are of the form
\begin{subequations}
\begin{align}
    \psi_{d_{xz}} =& \ket{\mathrm{Cu},\mathrm{X}, xz, 1}\begin{pmatrix}1 \\ 0\end{pmatrix} +e^{\frac{\pi}{4}i} \ket{\mathrm{Cu},X, yz, 2}\begin{pmatrix}0 \\ 1\end{pmatrix}, \\
    \psi_{d_{yz}} =& \ket{\mathrm{Cu},\mathrm{X}, yz, 1}\begin{pmatrix}1 \\ 0\end{pmatrix} +e^{\frac{\pi}{4}i} \ket{\mathrm{Cu},X, xz, 2}\begin{pmatrix}0 \\ 1\end{pmatrix},
    \\ \psi_{d_{xy}} =& \ket{\mathrm{Cu},\mathrm{X}, xy, 1}\begin{pmatrix}0 \\ 1\end{pmatrix} +e^{\frac{3\pi}{4}i} \ket{\mathrm{Cu},\mathrm{X}, xy, 2}\begin{pmatrix}1 \\ 0\end{pmatrix}.
\end{align}
\label{BasisfunctionsXCu}
\end{subequations}
The non-symmorphic symmetries exchange the locations $n=1,2$ along with a spin-flip. The X$_4$ basis functions have a similar form, except that the $n=2$ part has a relative minus sign. Evaluating the Pauli spin matrices with respect to the basis functions at the X-point reveals zero spin polarization in all directions, as consistent with the calculations shown in Figure \ref{fig:spinPolarization}.

The eigenvalues of the effective Hamiltonian in Eq. \eqref{hamXPoint} 
\begin{subequations}
\begin{align}
\begin{split}
    E_- =& E_2 + (A_2+C_2)(\kappa_x^2 + \kappa_y^2) + B_2\kappa_z^2\\ &+ 2(A_2-C_2)\kappa_x\kappa_y + \frac{\hbar^2}{2m}\kappa^2,
\end{split}\\
\begin{split}
    E_+ =& E_4 + (A_4+C_4)(\kappa_x^2 + \kappa_y^2) + B_4\kappa_z^2\\  &-2(A_4-C_4)\kappa_x\kappa_y + \frac{\hbar^2}{2m}\kappa^2,
\end{split}
\end{align}
\end{subequations}
illustrate the band dispersion at the valence band maximum to second order in $\bm{\kappa}$.

The effective mass tensor is defined as
\begin{align}
    (m^*_{ij})^{-1} = \frac{1}{\hbar^2}\frac{\partial^2 E}{\partial \kappa_i \partial \kappa_j}
\end{align}
and is, in general, valley specific. The effective mass tensor at a related valley is
\begin{align}
    m^*_{ij} = \mathcal{R}_{ii'}\mathcal{R}_{jj'} m_{i'j'}^*
    \label{effectiveMassRelation}
\end{align}
where $\mathcal{R}$ is a symmetry transformation relating the two valleys.
Now, for the two bands at the X-point, the effective masses differ in magnitude but have the same tensor form
\begin{align}
    m^* = \begin{pmatrix}
    m_{xx} & m_{xy} & 0 \\ m_{xy} & m_{xx} & 0 \\ 0 & 0 & m_{zz}
    \end{pmatrix}.
    \label{XMass}
\end{align}
From numerical calculations, we estimate that
\begin{subequations}
\begin{align}
    m^{(4)}_{xx}=-1.48, && m^{(4)}_{zz}=-1.36 , && m^{(4)}_{xy}=0.63,
\end{align}
\begin{align}
    m^{(2)}_{xx}=-1.40, && m^{(2)}_{zz}=-1.37 , && m^{(2)}_{xy}=0.70
\end{align}
\end{subequations}
in units of the bare electron mass $m$. The superscripts refer to the upper X$_4$-band and the lower X$_2$-band at the valence band maximum.

\subsection{\label{subsec:Npoint}Conduction band minimum}
In this section, we consider the conduction band minimum located at the N-point. The N-point has a low symmetry and a fourfold valley degeneracy. The valleys are at $\mathrm{N}=2\pi(1/2a, 0, 1/2c)$, $\mathrm{N}'=2\pi(-1/2a, 0, 1/2c)$, $\mathrm{N}''=2\pi(0, 1/2a, -1/2c)$ and $\mathrm{N}'''=2\pi(0, -1/2a, -1/2c)$. Each conduction band minimum consists of two non-degenerate bands. The lower band transforms as the N$_4$ irrep, whereas the upper band transforms as the N$_3$ irrep. The bands are split by 2.9 meV due to SOC. We perform a similar analysis as we did for the X-point, starting with the band coupling shown in Table \ref{Tab:MatrixProductsN}.
\begin{table}[ht]
\caption{Selection rules for electronic states at the N-point.}
\centering
\begin{ruledtabular}
\begin{tabular}{ p{8cm} }
\vspace{-7mm}
{\begin{align*}
 \frac{\hbar}{m}\bra{u^{N_3}}(P_x, P_y, P_z)\ket{u^{N_3}} &= (0, a_1 , 0)  \\
 \frac{\hbar}{m}\bra{u^{N_4}}(P_x, P_y, P_z)\ket{u^{N_4}} &= (0, a_2, 0) \\
 \frac{\hbar}{m}\bra{u^{N_3}}(P_x, P_y, P_z)\ket{u^{N_4}} &= (a_{34}, 0, b_{34})
  \end{align*}}
\vspace{-4mm}
\end{tabular}
\label{Tab:symmetries2}
\end{ruledtabular}
\label{Tab:MatrixProductsN}
\end{table}

The intra-band coupling to the first order in the crystal momentum deviation $\bm{\kappa}$ is again zero at band extrema. We find the two-band Hamiltonian
\begin{widetext}
\begin{subequations}
\begin{align}
    \mathcal{H}_{11}=&E_3  + A_3\kappa_y^2 + B_3\kappa_x^2 + C_3\kappa_z^2 + (R_3+R_3^*)\kappa_x\kappa_z + \frac{\hbar^2}{2m}\kappa^2,\\ 
    \label{offDiag1} \mathcal{H}_{12} =& a_{34}\kappa_x + b_{34}\kappa_z + (S_a + S_b) \kappa_x\kappa_y + (T_a + T_b) \kappa_y \kappa_z, \\ \mathcal{H}_{21} =& a_{34}^*\kappa_x + b_{34}^*\kappa_z +(S_a^* + S_b^*) \kappa_x\kappa_y + (T_a^* + T_b^*) \kappa_y \kappa_z \label{offDiag2}, \\ \mathcal{H}_{22} =& E_4  + A_4\kappa_y^2 + B_4\kappa_x^2 + C_4\kappa_z^2 + (R_4+R_4^*)\kappa_x\kappa_z + \frac{\hbar^2}{2m}\kappa^2.
\end{align}
\label{Nmodel}
\end{subequations}
\end{widetext}
Table \ref{Tab:MatrixProductsN} defines the first-order coupling constants $a_{34}$ and $b_{34}$ and the coupling parameters to second order are defined in Appendix \ref{sec:DefinitionOfFreeParameters}.2. The corresponding eigenvalues to second order in the crystal momentum deviation $\bm{\kappa}$ are
\begin{subequations}
\begin{align}
\begin{split}
    E_+ =& E_3+A_3\kappa_y^2+B_3\kappa_x^2 + C_3\kappa_z^2+ (R_3+R_3^*)\kappa_x\kappa_z\\ &+ \frac{\abs{a_{34}}^2\kappa_x^2 +\abs{b_{34}}^2\kappa_z^2 + (a_{34}b_{34}^* + a_{34}^*b_{34})\kappa_x\kappa_z}{E_3-E_4} \\ &+ \frac{\hbar^2}{m^2}\kappa^2,
\end{split}
\end{align}
\begin{align}
\begin{split}
    E_- =& E_4+A_4\kappa_y^2+B_4\kappa_x^2 + C_4\kappa_z^2 (R_4+R_4^*)\kappa_x\kappa_z\\  &-\frac{\abs{a_{34}}^2\kappa_x^2 +\abs{b_{34}}^2\kappa_z^2 + (a_{34}b_{34}^* + a_{34}^*b_{34})\kappa_x\kappa_z}{E_3-E_4} \\ &+ \frac{\hbar^2}{m^2}\kappa^2.
\end{split}
\end{align}
\label{Ndisp}
\end{subequations}
The second-order terms in Eqs. \eqref{offDiag1} and \eqref{offDiag2} do not contribute to the effective mass of the bands, but to the fourth order in momentum deviation $\bm{\kappa}$. We disregard these terms.

The wave functions constituting the conduction band minimum have large Fe $d_{xz}$ contributions. We now explore the N$_3$ and N$_4$ basis functions of this type. The basis function for the N$_3$ state has the form
\begin{align}
    \psi_{d_{xz}} = \ket{\mathrm{Fe},\mathrm{N}, xz, 1}\begin{pmatrix}\alpha \\ \beta\end{pmatrix}  +i\ket{\mathrm{Fe},\mathrm{N}, xz, 2}\begin{pmatrix}\beta \\ -\alpha\end{pmatrix}.
    \label{BasisfunctionsNFe}
\end{align}
The analogous basis function of the N$_4$ irrep has a relative minus sign for the $n=2$ part. The apparent $x$-$y$ asymmetry is compensated by the related valleys. The N-point has a low symmetry. Hence, the spinor part of the basis functions has a more general form than for the X-point. Evaluating the spin Pauli matrices with respect to the basis functions, we find a finite spin polarization in the $y$-direction consistent with Figure \ref{fig:spinPolarization} (b). 

Based on the dispersion relation of the two N-bands in Eq. \eqref{Ndisp}, the effective mass tensor is of the form
\begin{align}
    m^* = \begin{pmatrix}
    m_{xx} & 0 & m_{xz} \\ 0 & m_{yy} & 0 \\ m_{xz} & 0 & m_{zz}
    \end{pmatrix}.
    \label{eq:Nmass}
\end{align}
We find the numerical estimates of effective mass tensors to be
\begin{subequations}
\begin{align}
    m^{(4)}_{xx}=0.72, && m^{(4)}_{yy} = 0.84, \\ m^{(4)}_{zz}=0.88, && m^{(4)}_{xz} = 0.89,
\end{align}
\end{subequations}
\begin{subequations}
\begin{align}
    m^{(3)}_{xx}=0.86, && m^{(3)}_{yy} = 0.83, \\ m^{(3)}_{zz}=0.89, && m^{(3)}_{xz} = 0.92.
\end{align}
\end{subequations}
in units of the bare electron mass $m$. The superscripts refer to the lower N$_4$ state and the upper N$_3$ state constituting the CBM.

\section{\label{sec:External magnetic field}External magnetic field}

In this section, we extend the effective model for the valence electrons at the X-point and the conduction electrons at the N-point to include the effects of an external magnetic field.

We employ the Kohn-Luttinger transcription $\bm{\kappa}\rightarrow (-i\bm{\nabla} - e\bm{A})$ and account for the external magnetic field through the vector potential $\bm{A}$. This yields a system of envelope function differential equations
\begin{align}
\begin{split}
    \sum_{n'}\left[ D_{nn'ij}(-i\nabla_i-eA_i)(-i\nabla_j-eA_j)\right]F_{n'}(\bm{r})\\=EF_{n}(\bm{r}),
\end{split}
\end{align}
where $F_n$ are envelope functions. Products of non-commuting factors should be interpreted as symmetrized products \cite{luttinger1955motion, luttinger1956quantum}. To that end, we consider both the symmetric and antisymmetric terms
\begin{align}
    D_{nn'ij}\kappa_{i}\kappa_{j} = \frac{1}{2}D_{nn'ij}^{(S)}\left\{\kappa_{i}, \kappa_{j}\right\} + \frac{1}{2}D_{nn'ij}^{(A)}\left[\kappa_{i}, \kappa_{j}\right],
\end{align}
Here, the $\left\{\kappa_{i}, \kappa_{j}\right\}= \kappa_i \kappa_j + \kappa_j \kappa_i$ is the anticommutator and the commutator
\begin{align}
    \left[\kappa_{i}, \kappa_{j}\right] = \frac{ie}{\hbar c}B_k
\end{align}
is finite in the presence of an external magnetic field $B_{k}$. The indices $i$, $j$, and $k$ form a right-handed Cartesian coordinate system. The symmetric and antisymmetric terms are defined as
\begin{subequations}
\begin{align}
    D_{nn'ij}^{(S)} = \frac{1}{2}\left(D_{nn'ij} + D_{nn'ji}\right), \\
    D_{nn'ij}^{(A)} = \frac{1}{2}\left(D_{nn'ij} - D_{nn'ji}\right),
\end{align}
\end{subequations}
with
\begin{align}
    D_{nn'ij} = \frac{\hbar^2}{2m^2}\sum_{n''} \frac{\bra{n}P_{i}\ket{n''}\bra{n''}P_{j}\ket{n'}}{E_n - E_{n''}}.
\end{align}
The symmetric terms are identical to the coefficients in Eq. \eqref{Xmodel} and \eqref{Nmodel}. The antisymmetric terms transform as the Zeeman coupling $H_{\mathrm{Z}}=\mu_B \bm{\sigma}\cdot\bm{B}$ and give rise to an effective coupling constant $g_{\mathrm{eff}}$. In general, we find that the effective Zeeman coupling tensor takes a different form than for conventional non-magnetic semiconductors. This can be understood from the form of the basis functions in Eqs. \eqref{BasisfunctionsXS}, \eqref{BasisfunctionsXCu} and \eqref{BasisfunctionsNFe}.

For the X-point, we find
\begin{subequations}
\begin{align}
    D_{12}^{(A)} =& \mu_B \bra{X_2}\sigma_z\ket{X_4} B_z - (M-L) \frac{ie}{\hbar c} B_z, \\
    D_{21}^{(A)}=& \mu_B \bra{X_4}\sigma_z\ket{X_2} B_z - (M^*-L^*)\frac{ie}{\hbar c} B_z,
\end{align}
\end{subequations}
where the first terms are the spin Zeeman effect and the second terms arise from coupling to external bands.
Note that an external magnetic field in the $xy$-plane does not couple directly to the electronic states at the X-point within the regime of the two-band model. This is consistent with the form of the basis functions at the VBM.

Next, we consider the Zeeman effect and the antisymmetric terms for the N-point. We find
\begin{subequations}
\begin{align}
    D_{11}^{(A)} =& \mu_B \bra{N_3}\sigma_y\ket{N_3} B_y - (R_3-R_3^*)\frac{ie}{\hbar c} B_y, \\
    \begin{split}
    D_{12}^{(A)} =& \mu_B \bra{N_3}\sigma_x\ket{N_4} B_x + \mu_B \bra{N_3}\sigma_z\ket{N_4} B_{z} \\&- (S_a - S_b)\frac{ie}{\hbar c}B_z + (T_a - T_b)\frac{ie}{\hbar c}B_x,
    \end{split}\\
    \begin{split}
    D_{21}^{(A)} =& \mu_B \bra{N_4}\sigma_x\ket{N_3} B_x + \mu_B \bra{N_4}\sigma_z\ket{N_3} B_z \\&- (S_a^* - S_b^*)\frac{ie}{\hbar c}B_z + (T_a^* - T_b^*)\frac{ie}{\hbar c}B_x,
    \end{split}\\
    D_{22}^{(A)} =&\mu_B \bra{N_4}\sigma_y\ket{N_4} B_y - (R_4 - R_4^*)\frac{ie}{\hbar c} B_y.
\end{align}
\end{subequations}
The N-point has a lower symmetry than the X-point. Here, the bands are split by an external magnetic field in all directions. There is a finite Zeeman coupling for each of the bands individually in the $y$-direction. Figure \ref{fig:spinPolarization} (b) shows that $\bra{N_3}\sigma_y\ket{N_3} \approx -\bra{N_4}\sigma_y\ket{N_4}$ at the CBM.

\section{\label{sec:Cyclotron resonance}Cyclotron resonance}
The values of the effective mass tensors of the valence band maxima and the conduction band minima can be experimentally determined. One possibility is to measure the cyclotron resonance due to an AC electric field $\bm{E}$ in the presence of a static magnetic field $\bm{B}$. In this section, we investigate how the cyclotron resonance depends on the direction of the magnetic field. As discussed in section \ref{sec:Ab-initio calculations} B, both the VBM and CBM have a two-band structure induced by SOC. As shown in section \ref{sec:Effective theory}, the effective masses of these split bands differ. Hence, this structure may give rise to twice the number of resonance peaks as for a single band. The onset of the extra peaks should depend on temperature and carrier density. Consider the equation of motion
\begin{align}
    \frac{\mathrm{d}(m^*\bm{v}_{\mathrm{d}})}{\mathrm{d}t} + \frac{m^*\bm{v}_{\mathrm{d}}}{\tau_m} = e(\bm{E}+\left[\bm{v}_{\mathrm{d}}\times\bm{B}\right]),
    \label{eq:equationOfMotion}
\end{align}
where $\bm{v}_\mathrm{d}$ is the carrier drift velocity, $m^*$ is the effective mass and $\tau_m$ is the scattering time \cite{seeger2013semiconductor}. To solve Eq. \eqref{eq:equationOfMotion} within the many-valley model of chalcopyrite, we introduce a coordinate system $(\bm{\hat{e}}_1,\bm{\hat{e}}_2,\bm{\hat{e}}_3)$ that diagonalizes the effective mass tensor. Furthermore, we define directional cosines of the magnetic field $\bm{B}$ with respect to this coordinate system as
\begin{align}
    \alpha = \frac{\bm{B}\cdot\bm{\hat{e}}_1}{\abs{\bm{B}}} && \beta = \frac{\bm{B}\cdot\bm{\hat{e}}_2}{\abs{\bm{B}}} && \gamma = \frac{\bm{B}\cdot\bm{\hat{e}}_3}{\abs{\bm{B}}}.
\end{align}
These coordinates diagonalize the effective mass tensor such that
\begin{align}
    m^* = \begin{pmatrix} m_1 & 0 & 0 \\ 0 & m_2 & 0 \\ 0 & 0 & m_3 \end{pmatrix}.
\end{align}
Eq. \eqref{eq:equationOfMotion} yields $\omega_{\mathrm{c}} = (e/m^*)B$ for the resonance frequency, where
\begin{align}
    m^* = \sqrt{\frac{m_1 m_2 m_3}{\alpha^2 m_1 + \beta^2 m_2 + \gamma^2 m_3}}.
    \label{eq:effectiveMass}
\end{align}
Here, we neglected $\bm{E}$ and $\tau_m^{-1}$ at the resonance frequency for simplicity. Both the diagonalizing coordinate system and the effective mass tensor are, in general, valley-specific. The effective mass tensors at related valleys are found from Eq. \eqref{effectiveMassRelation}.

Now we consider Eq. \eqref{eq:effectiveMass} for the VBM explicitly. The effective mass tensor in Eq. \eqref{XMass} is diagonal in the basis
\begin{align}
    \bm{\hat{e}}_1 = \frac{\sqrt{2}}{2}(1,1,0), && \bm{\hat{e}}_2 = \frac{\sqrt{2}}{2}(-1,1,0), && \bm{\hat{e}}_3 = (0,0,1),
\end{align}
with principle effective masses
\begin{align}
    m_1^X = m_{xx}+m_{xy}, && m_2^X = m_{xx}-m_{xy},  && m_3^X = m_{zz},
\end{align}
at the $\mathrm{X}$-point. The related effective masses at $\mathrm{X'}$ are
\begin{align}
    m_1^{X'} = m_2^X, && m_2^{X'} = m_1^{X}, && m_3^{X'} = m_3^X.
\end{align}
The two valleys are equivalent with respect to a magnetic field in the $\hat{z}$-direction. The effective mass in such a configuration is
\begin{align}
    m^* = \sqrt{m_1m_2} = \sqrt{m_{xx}^2 - m_{xy}^2}.
\end{align}
For a magnetic field, $B$ in the $xy$-plane, with a polar angle $\theta$ with respect to the $x$-axis, the effective mass is
\begin{align}
    m^* = \sqrt{\frac{m_1m_2m_3}{m_1\sin^2{\theta} + m_2\cos^2{\theta}}}.
\end{align}
Hence, in general, the two valleys give two distinct cyclotron resonance frequencies for an in-plane magnetic field. The number of- and the relation between the cyclotron resonance frequencies can be used to verify that the VBM is located at the X-point.

The CBMs consist of the four inequivalent $\mathrm{N}$-valleys as defined in section \ref{subsec:Npoint}. Each valley has a distinct effective mass tensor, although the tensors are related as in Eq. \eqref{effectiveMassRelation}. The coordinate system in which each tensor is diagonal differs. The four sets of unit vectors are

\begin{align}
\begin{split}
    \bm{\hat{e}}_1 &= (0,1,0), \\
    \bm{\hat{e}}_2 &= C_-\left(\frac{m_{xx} - m_{zz} - A}{2m_{xz}}, 0,1\right), \\
    \bm{\hat{e}}_3 &= C_+\left(\frac{m_{xx} - m_{zz} + A}{2m_{xz}}, 0,1\right),
\end{split}
\end{align}
for the $\mathrm{N}$-point,
\begin{align}
\begin{split}
    \bm{\hat{e}}_1' &= (0,1,0), \\
    \bm{\hat{e}}_2' &= C_-\left(\frac{-m_{xx} + m_{zz} + A}{2m_{xz}}, 0,1\right), \\
    \bm{\hat{e}}_3' &= C_+\left(\frac{-m_{xx} + m_{zz} - A}{2m_{xz}}, 0,1\right).
\end{split}
\end{align}
for the $\mathrm{N'}$-point,
\begin{align}
\begin{split}
    \bm{\hat{e}}_1'' &= (1,0,0), \\
    \bm{\hat{e}}_2'' &= C_-\left(0, \frac{-m_{xx} + m_{zz} + A}{2m_{xz}}, 0,1\right), \\
    \bm{\hat{e}}_3'' &= C_+\left(0,\frac{-m_{xx} + m_{zz} - A}{2m_{xz}},1\right),
    \label{eq:N''Coordinates}
\end{split}
\end{align}
for the $\mathrm{N''}$-point and
\begin{align}
\begin{split}
    \bm{\hat{e}}_1''' &= (1,0,0), \\
    \bm{\hat{e}}_2''' &= C_-\left(0, \frac{m_{xx} - m_{zz} - A}{2m_{xz}}, 0,1\right), \\
    \bm{\hat{e}}_3''' &= C_+\left(0,\frac{m_{xx} - m_{zz} + A}{2m_{xz}},1\right),
    \label{eq:N'''Coordinates}
\end{split}
\end{align}
for the $\mathrm{N'''}$-point. We introduced the variable $A=\sqrt{(m_{xx}-m_{zz})^2 + 4m_{xz}^2}$ for notational convenience. The normalization constants are
\begin{subequations}
\begin{align}
    C_+ = \left(\frac{1}{2}\sqrt{4+\left(\frac{m_{xx}-m_{zz} + A}{2m_{xz}}\right)^2}\right)^{-1}, \\
    C_- = \left(\frac{1}{2}\sqrt{4+\left(\frac{m_{xx}-m_{zz} - A}{2m_{xz}}\right)^2}\right)^{-1}.
\end{align}
\end{subequations}
The effective mass tensor for each valley is diagonal with respect to their coordinate system. For each of the four valleys, the diagonal elements are
\begin{subequations}
\begin{align}
    m_1 =& m_{yy}, \\ m_2 =& \frac{1}{2}\left(m_{xx} + m_{zz} - A\right), \\ m_3 =& \frac{1}{2}\left(m_{xx} + m_{zz} + A\right).
\end{align}
\end{subequations}
In this way, cyclotron resonance is a good way to verify the suggested multivalley structure of both the VBM and CBM.

\section{\label{sec:Conclusions}Conclusions}
CuFeS$_2$ is a semiconducting collinear antiferromagnet with a non-symmorphic crystal lattice. Its magnetic space group allows for intriguing properties such as spin-polarized electron bands and the anomalous Hall effect. We have explored its low-energy electron properties based on its magnetic symmetry group and DFT calculations. On phenomenological grounds, we found that the conductivity has components scaling linearly with the Néel vector. This included both longitudinal terms and terms corresponding to the anomalous Hall effect. The electron dispersion rendered a multivalley semiconductor with an indirect band gap of $0.779$ eV. Consistent with the magnetic symmetry class, we found the electron bands to be partly spin polarized. In particular, we found an in-plane spin polarization at the conduction band minimum. We developed effective $\bm{k}\cdot\bm{p}$ models of the VBM and CBM. The resulting effective mass tensors were quantified by the \textit{ab initio} calculations. We extended the models to include an external magnetic field using the envelope function approximation. Lastly, we suggested how to verify the effective mass tensors experimentally. The approach takes into account the multivalley band structure and serves as a framework for measuring the effective mass at and verifying the location of the determined CBM and VBM, specifically.

\begin{acknowledgments}
The Research Council of Norway supported this work through its Centres of Excellence funding scheme, project number 262633, "QuSpin". The Norwegian Metacenter for Computational Science provided computational resources at Uninett Sigma 2, project number NN9301K.
\end{acknowledgments}

\appendix

\section{Evaluation of matrix products}

We evaluate the matrix products based on the symmetry of the respective wave functions and the transformation properties of operators. As an example, we consider the matrix product
\begin{align}
    \bra{X} \hat{O} \ket{Y},
\end{align}
where the bra state, ket state, and the operator transform as the irreducible representations $\Gamma_X$ and $\Gamma_Y$, $\Gamma_{\hat{O}}$, respectively. To determine if the matrix product is finite, we consider the corresponding tensor product
\begin{align}
    \Gamma_X \otimes \Gamma_{\hat{O}} \otimes \Gamma_Y = \bigoplus_i \Gamma_i,
\end{align}
where the right-hand side is a direct sum of irreducible representations. The matrix product is then finite if and only if the sum on the right-hand side contains the trivial representation $\Gamma_I$. Furthermore, we can relate distinct matrix products by using that the value of all matrix products is invariant under the relevant symmetry operations.

\section{\label{sec:DefinitionOfFreeParameters}Definition of free parameters}

In the $\bm{k}\cdot\bm{p}$ theory, we consider the coupling between bands to second order in the $\mathcal{H}_{\bm{\kappa\cdot\bm{p}}}$ perturbation in Eq. \eqref{Hkp}. The resulting free parameters are typically treated semi-empirically. In this work, we consider them as free parameters to be determined experimentally. In the following, we define the free parameters used in section \ref{sec:Effective theory} and \ref{sec:External magnetic field}.
\subsection{The X-point}
The X-point exhibits four distinct one-dimensional representations. We call these X$_2$, X$_3$, X$_4$ and X$_5$ consistent with the Bilbao Crystallographic Server \cite{elcoro2017double}.

First, we present the parameters describing the coupling between the X$_2$ band to all external bands
\begin{align}
    A_2 & = \frac{\hbar^2}{m^2}\sum_{i \in {\mathrm{Ir}}_3} \frac{\abs{\bra{X_2}P_x\ket{i}}^2}{E_2-E_i} = \frac{\hbar^2}{m^2}\sum_{i \in {\mathrm{Ir}}_3} \frac{\abs{\bra{X_2}P_y\ket{i}}^2}{E_2-E_i},
    \\
    B_2 & = \frac{\hbar^2}{m^2}\sum_{i \in {\mathrm{Ir}}_2} \frac{\abs{\bra{X_2}P_z\ket{i}}^2}{E_2-E_i}, \\
    C_2 & = \frac{\hbar^2}{m^2}\sum_{i \in {\mathrm{Ir}}_5} \frac{\abs{\bra{X_2}P_x\ket{i}}^2}{E_2-E_i} = \frac{\hbar^2}{m^2}\sum_{i \in {\mathrm{Ir}}_5} \frac{\abs{\bra{X_2}P_y\ket{i}}^2}{E_2-E_i}.
\end{align}

Here, the sums run over all bands $i$, which transform as the representation $\mathrm{Ir}_n$.
In addition, we include the processes mixing $P_x$, $P_y$ and $P_z$

\begin{align}
    A_2 = \frac{\hbar^2}{m^2}\sum_{i \in {\mathrm{Ir}}_3} \frac{\bra{X_2}P_x\ket{i}\bra{i}P_y\ket{X_2}}{E_2-E_i}, \\ A_2 = \frac{\hbar^2}{m^2}\sum_{i \in {\mathrm{Ir}}_3} \frac{\bra{X_2}P_y\ket{i}\bra{i}P_x\ket{X_2}}{E_2-E_i},\\-C_2 = \frac{\hbar^2}{m^2}\sum_{i \in {\mathrm{Ir}}_5} \frac{\bra{X_2}P_x\ket{i}\bra{i}P_y\ket{X_2}}{E_2-E_i}, \\ -C_2 =\frac{\hbar^2}{m^2}\sum_{i \in {\mathrm{Ir}}_5} \frac{\bra{X_2}P_y\ket{i}\bra{i}P_x\ket{X_2}}{E_2-E_i}.
\end{align}
Note how the non-symmorphic mirror symmetries $(\sigma_{xy}\vert \bm{\tau})$ and  $(\sigma_{\bar{x}y}\vert \bm{\tau})$ relate the $P_x$ and $P_y$ components.

Now we consider the coupling of the X$_4$ band to all other bands
\begin{align}
    A_4 & = \frac{\hbar^2}{m^2}\sum_{i \in {\mathrm{Ir}}_3} \frac{\abs{\bra{X_4}P_x\ket{i}}^2}{E_4-E_i} = \frac{\hbar^2}{m^2}\sum_{i \in {\mathrm{Ir}}_3} \frac{\abs{\bra{X_4}P_y\ket{i}}^2}{E_4-E_i},\\
    B_4 & = \frac{\hbar^2}{m^2}\sum_{i \in {\mathrm{Ir}}_4} \frac{\abs{\bra{X_4}P_z\ket{i}}^2}{E_4-E_i}, \\
    C_4 & = \frac{\hbar^2}{m^2}\sum_{i \in {\mathrm{Ir}}_5} \frac{\abs{\bra{X_4}P_x\ket{i}}^2}{E_4-E_i} = \frac{\hbar^2}{m^2}\sum_{i \in {\mathrm{Ir}}_5} \frac{\abs{\bra{X_4}P_y\ket{i}}^2}{E_4-E_i}.
\end{align}

The mixed-momentum components are
\begin{align}
    -A_4 = \frac{\hbar^2}{m^2}\sum_{i \in {\mathrm{Ir}}_3} \frac{\bra{X_4}P_x\ket{i}\bra{i}P_y\ket{X_4}}{E_4-E_i}, \\ -A_4=\frac{\hbar^2}{m^2}\sum_{i \in {\mathrm{Ir}}_3} \frac{\bra{X_4}P_y\ket{i}\bra{i}P_x\ket{X_4}}{E_4-E_i},\\ C_4 = \frac{\hbar^2}{m^2}\sum_{i \in {\mathrm{Ir}}_5} \frac{\bra{X_4}P_x\ket{i}\bra{i}P_y\ket{X_4}}{E_4-E_i}, \\ C_4 = \frac{\hbar^2}{m^2}\sum_{i \in {\mathrm{Ir}}_5} \frac{\bra{X_4}P_y\ket{i}\bra{i}P_x\ket{X_4}}{E_4-E_i}.
\end{align}

Now we consider the off-diagonal terms. That is, coupling between the X$_2$ and X$_4$ band to second order in momentum
\begin{align}
    M = \frac{\hbar^2}{m^2}\sum_{i \in {\mathrm{Ir}}_3} \frac{\bra{X_2}P_x\ket{i}\bra{i}P_x\ket{X_4}}{E_4-E_i},  \\ -M=\frac{\hbar^2}{m^2}\sum_{i \in {\mathrm{Ir}}_2} \frac{\bra{X_2}P_y\ket{i}\bra{i}P_y\ket{X_4}}{E_4-E_i},\\
    L = \frac{\hbar^2}{m^2}\sum_{i \in {\mathrm{Ir}}_5} \frac{\bra{X_2}P_x\ket{i}\bra{i}P_x\ket{X_4}}{E_4-E_i}, \\ -L = \frac{\hbar^2}{m^2}\sum_{i \in {\mathrm{Ir}}_5} \frac{\bra{X_2}P_y\ket{i}\bra{i}P_y\ket{X_4}}{E_4-E_i}.
\end{align}
Now we consider mixing of $P_x$, $P_y$ and $P_z$ terms
\begin{align}
    -M = \frac{\hbar^2}{m^2}\sum_{i \in {\mathrm{Ir}}_3} \frac{\bra{X_2}P_x\ket{i}\bra{i}P_y\ket{X_4}}{E_4-E_i}, \\ M = \frac{\hbar^2}{m^2}\sum_{i \in {\mathrm{Ir}}_3} \frac{\bra{X_2}P_y\ket{i}\bra{i}P_x\ket{X_4}}{E_4-E_i},\\
    L = \frac{\hbar^2}{m^2}\sum_{i \in {\mathrm{Ir}}_5} \frac{\bra{X_2}P_x\ket{i}\bra{i}P_y\ket{X_4}}{E_4-E_i}, \\ -L = \frac{\hbar^2}{m^2}\sum_{i \in {\mathrm{Ir}}_5} \frac{\bra{X_2}P_y\ket{i}\bra{i}P_x\ket{X_4}}{E_4-E_i}.
\end{align}

\subsection{\label{sec:DefinitionOfFreeParametersN}The N-point}
In this section, we define the free parameters related to the effective description of the conduction band minimum. For each of the two bands, these are
\begin{align}
    A_3 = \frac{\hbar^2}{m^2}\sum_{i \in {\mathrm{Ir}}_3} \frac{\abs{\bra{N_3}P_y\ket{i}}^2}{E_3-E_i}, \\  A_4= \frac{\hbar^2}{m^2}\sum_{i \in {\mathrm{Ir}}_4} \frac{\abs{\bra{N_4}P_y\ket{i}}^2}{E_4-E_i},\\ B_3 = \frac{\hbar^2}{m^2}\sum_{i \in {\mathrm{Ir}}_4} \frac{\abs{\bra{N_4}P_x\ket{i}}^2}{E_4-E_i}, \\ B_4 = \frac{\hbar^2}{m^2}\sum_{i \in {\mathrm{Ir}}_3} \frac{\abs{\bra{N_3}P_x\ket{i}}^2}{E_3-E_i}, \\ C_3=\frac{\hbar^2}{m^2}\sum_{i \in {\mathrm{Ir}}_4} \frac{\abs{\bra{N_3}P_z\ket{i}}^2}{E_3-E_i}, \\  C_4 =\frac{\hbar^2}{m^2}\sum_{i \in {\mathrm{Ir}}_3} \frac{\abs{\bra{N_4}P_z\ket{i}}^2}{E_4-E_i}
    \\ R_3 = \frac{\hbar^2}{m^2}\sum_{i \in {\mathrm{Ir}}_4} \frac{\bra{N_3}P_x\ket{i}\bra{i}P_z\ket{N_3}}{E_3-E_i}, \\
    R_4 = \frac{\hbar^2}{m^2}\sum_{i \in {\mathrm{Ir}}_3} \frac{\bra{N_4}P_x\ket{i}\bra{i}P_z\ket{N_4}}{E_4-E_i}.
\end{align}
The off-diagonal coupling is captured by the free parameters
\begin{align}
    S_a=\frac{\hbar^2}{m^2}\sum_{i \in {\mathrm{Ir}}_{3}} \frac{\bra{N_3}P_y\ket{i}\bra{i}P_x\ket{N_4}}{E_3-E_i}, \\
    S_b=\frac{\hbar^2}{m^2}\sum_{i \in {\mathrm{Ir}}_{4}} \frac{\bra{N_3}P_x\ket{i}\bra{i}P_y\ket{N_4}}{E_3-E_i},\\
    T_a=\frac{\hbar^2}{m^2}\sum_{i \in {\mathrm{Ir}}_{3}} \frac{\bra{N_3}P_y\ket{i}\bra{i}P_z\ket{N_4}}{E_3-E_i}, \\
    T_b=\frac{\hbar^2}{m^2}\sum_{i \in {\mathrm{Ir}}_{4}} \frac{\bra{N_3}P_z\ket{i}\bra{i}P_y\ket{N_4}}{E_3-E_i}.
\end{align}


\bibliography{apssamp}

\providecommand{\noopsort}[1]{}\providecommand{\singleletter}[1]{#1}%
\begin{thebibliography}{51}%
\makeatletter
\providecommand \@ifxundefined [1]{%
 \@ifx{#1\undefined}
}%
\providecommand \@ifnum [1]{%
 \ifnum #1\expandafter \@firstoftwo
 \else \expandafter \@secondoftwo
 \fi
}%
\providecommand \@ifx [1]{%
 \ifx #1\expandafter \@firstoftwo
 \else \expandafter \@secondoftwo
 \fi
}%
\providecommand \natexlab [1]{#1}%
\providecommand \enquote  [1]{``#1''}%
\providecommand \bibnamefont  [1]{#1}%
\providecommand \bibfnamefont [1]{#1}%
\providecommand \citenamefont [1]{#1}%
\providecommand \href@noop [0]{\@secondoftwo}%
\providecommand \href [0]{\begingroup \@sanitize@url \@href}%
\providecommand \@href[1]{\@@startlink{#1}\@@href}%
\providecommand \@@href[1]{\endgroup#1\@@endlink}%
\providecommand \@sanitize@url [0]{\catcode `\\12\catcode `\$12\catcode
  `\&12\catcode `\#12\catcode `\^12\catcode `\_12\catcode `\%12\relax}%
\providecommand \@@startlink[1]{}%
\providecommand \@@endlink[0]{}%
\providecommand \url  [0]{\begingroup\@sanitize@url \@url }%
\providecommand \@url [1]{\endgroup\@href {#1}{\urlprefix }}%
\providecommand \urlprefix  [0]{URL }%
\providecommand \Eprint [0]{\href }%
\providecommand \doibase [0]{https://doi.org/}%
\providecommand \selectlanguage [0]{\@gobble}%
\providecommand \bibinfo  [0]{\@secondoftwo}%
\providecommand \bibfield  [0]{\@secondoftwo}%
\providecommand \translation [1]{[#1]}%
\providecommand \BibitemOpen [0]{}%
\providecommand \bibitemStop [0]{}%
\providecommand \bibitemNoStop [0]{.\EOS\space}%
\providecommand \EOS [0]{\spacefactor3000\relax}%
\providecommand \BibitemShut  [1]{\csname bibitem#1\endcsname}%
\let\auto@bib@innerbib\@empty
\bibitem [{\citenamefont {Jungwirth}\ \emph {et~al.}(2016)\citenamefont
  {Jungwirth}, \citenamefont {Marti}, \citenamefont {Wadley},\ and\
  \citenamefont {Wunderlich}}]{Jungwirth:NNano2016}%
  \BibitemOpen
  \bibfield  {author} {\bibinfo {author} {\bibfnamefont {T.}~\bibnamefont
  {Jungwirth}}, \bibinfo {author} {\bibfnamefont {X.}~\bibnamefont {Marti}},
  \bibinfo {author} {\bibfnamefont {P.}~\bibnamefont {Wadley}},\ and\ \bibinfo
  {author} {\bibfnamefont {J.}~\bibnamefont {Wunderlich}},\ }\bibfield  {title}
  {\bibinfo {title} {Antiferromagnetic spintronics},\ }\href
  {https://doi.org/10.1038/nnano.2016.18} {\bibfield  {journal} {\bibinfo
  {journal} {Nat Nano}\ }\textbf {\bibinfo {volume} {11}},\ \bibinfo {pages}
  {231} (\bibinfo {year} {2016})}\BibitemShut {NoStop}%
\bibitem [{\citenamefont {Baltz}\ \emph {et~al.}(2018)\citenamefont {Baltz},
  \citenamefont {Manchon}, \citenamefont {Tsoi}, \citenamefont {Moriyama},
  \citenamefont {Ono},\ and\ \citenamefont {Tserkovnyak}}]{Baltz:RMP2018}%
  \BibitemOpen
  \bibfield  {author} {\bibinfo {author} {\bibfnamefont {V.}~\bibnamefont
  {Baltz}}, \bibinfo {author} {\bibfnamefont {A.}~\bibnamefont {Manchon}},
  \bibinfo {author} {\bibfnamefont {M.}~\bibnamefont {Tsoi}}, \bibinfo {author}
  {\bibfnamefont {T.}~\bibnamefont {Moriyama}}, \bibinfo {author}
  {\bibfnamefont {T.}~\bibnamefont {Ono}},\ and\ \bibinfo {author}
  {\bibfnamefont {Y.}~\bibnamefont {Tserkovnyak}},\ }\bibfield  {title}
  {\bibinfo {title} {Antiferromagnetic spintronics},\ }\href
  {https://doi.org/10.1103/RevModPhys.90.015005} {\bibfield  {journal}
  {\bibinfo  {journal} {Reviews of Modern Physics}\ }\textbf {\bibinfo {volume}
  {90}},\ \bibinfo {pages} {015005} (\bibinfo {year} {2018})}\BibitemShut
  {NoStop}%
\bibitem [{\citenamefont {Gomonay}\ \emph {et~al.}(2018)\citenamefont
  {Gomonay}, \citenamefont {Baltz}, \citenamefont {Brataas},\ and\
  \citenamefont {Tserkovnyak}}]{Gomonay:NPHYS2018}%
  \BibitemOpen
  \bibfield  {author} {\bibinfo {author} {\bibfnamefont {O.}~\bibnamefont
  {Gomonay}}, \bibinfo {author} {\bibfnamefont {V.}~\bibnamefont {Baltz}},
  \bibinfo {author} {\bibfnamefont {A.}~\bibnamefont {Brataas}},\ and\ \bibinfo
  {author} {\bibfnamefont {Y.}~\bibnamefont {Tserkovnyak}},\ }\bibfield
  {title} {\bibinfo {title} {Antiferromagnetic spin textures and dynamics},\
  }\href {https://doi.org/10.1038/s41567-018-0049-4} {\bibfield  {journal}
  {\bibinfo  {journal} {Nature Physics}\ }\textbf {\bibinfo {volume} {14}},\
  \bibinfo {pages} {213} (\bibinfo {year} {2018})}\BibitemShut {NoStop}%
\bibitem [{\citenamefont {Wadley}\ \emph {et~al.}(2016)\citenamefont {Wadley},
  \citenamefont {Howells}, \citenamefont {Železný}, \citenamefont {Andrews},
  \citenamefont {Hills}, \citenamefont {Campion}, \citenamefont {Novák},
  \citenamefont {Olejník}, \citenamefont {Maccherozzi}, \citenamefont {Dhesi},
  \citenamefont {Martin}, \citenamefont {Wagner}, \citenamefont {Wunderlich},
  \citenamefont {Freimuth}, \citenamefont {Mokrousov}, \citenamefont {Kuneš},
  \citenamefont {Chauhan}, \citenamefont {Grzybowski}, \citenamefont
  {Rushforth}, \citenamefont {Edmonds}, \citenamefont {Gallagher},\ and\
  \citenamefont {Jungwirth}}]{Wadley:SCIENCE2016}%
  \BibitemOpen
  \bibfield  {author} {\bibinfo {author} {\bibfnamefont {P.}~\bibnamefont
  {Wadley}}, \bibinfo {author} {\bibfnamefont {B.}~\bibnamefont {Howells}},
  \bibinfo {author} {\bibfnamefont {J.}~\bibnamefont {Železný}}, \bibinfo
  {author} {\bibfnamefont {C.}~\bibnamefont {Andrews}}, \bibinfo {author}
  {\bibfnamefont {V.}~\bibnamefont {Hills}}, \bibinfo {author} {\bibfnamefont
  {R.~P.}\ \bibnamefont {Campion}}, \bibinfo {author} {\bibfnamefont
  {V.}~\bibnamefont {Novák}}, \bibinfo {author} {\bibfnamefont
  {K.}~\bibnamefont {Olejník}}, \bibinfo {author} {\bibfnamefont
  {F.}~\bibnamefont {Maccherozzi}}, \bibinfo {author} {\bibfnamefont {S.~S.}\
  \bibnamefont {Dhesi}}, \bibinfo {author} {\bibfnamefont {S.~Y.}\ \bibnamefont
  {Martin}}, \bibinfo {author} {\bibfnamefont {T.}~\bibnamefont {Wagner}},
  \bibinfo {author} {\bibfnamefont {J.}~\bibnamefont {Wunderlich}}, \bibinfo
  {author} {\bibfnamefont {F.}~\bibnamefont {Freimuth}}, \bibinfo {author}
  {\bibfnamefont {Y.}~\bibnamefont {Mokrousov}}, \bibinfo {author}
  {\bibfnamefont {J.}~\bibnamefont {Kuneš}}, \bibinfo {author} {\bibfnamefont
  {J.~S.}\ \bibnamefont {Chauhan}}, \bibinfo {author} {\bibfnamefont {M.~J.}\
  \bibnamefont {Grzybowski}}, \bibinfo {author} {\bibfnamefont {A.~W.}\
  \bibnamefont {Rushforth}}, \bibinfo {author} {\bibfnamefont {K.~W.}\
  \bibnamefont {Edmonds}}, \bibinfo {author} {\bibfnamefont {B.~L.}\
  \bibnamefont {Gallagher}},\ and\ \bibinfo {author} {\bibfnamefont
  {T.}~\bibnamefont {Jungwirth}},\ }\bibfield  {title} {\bibinfo {title}
  {Electrical switching of an antiferromagnet},\ }\href
  {http://science.sciencemag.org/sci/early/2016/01/13/science.aab1031.full.pdf}
  {\bibfield  {journal} {\bibinfo  {journal} {Science}\ } (\bibinfo {year}
  {2016})}\BibitemShut {NoStop}%
\bibitem [{\citenamefont {Bodnar}\ \emph {et~al.}(2018)\citenamefont {Bodnar},
  \citenamefont {Šmejkal}, \citenamefont {Turek}, \citenamefont {Jungwirth},
  \citenamefont {Gomonay}, \citenamefont {Sinova}, \citenamefont {Sapozhnik},
  \citenamefont {Elmers}, \citenamefont {Kläui},\ and\ \citenamefont
  {Jourdan}}]{Bodnar:NatC2018}%
  \BibitemOpen
  \bibfield  {author} {\bibinfo {author} {\bibfnamefont {S.~Y.}\ \bibnamefont
  {Bodnar}}, \bibinfo {author} {\bibfnamefont {L.}~\bibnamefont {Šmejkal}},
  \bibinfo {author} {\bibfnamefont {I.}~\bibnamefont {Turek}}, \bibinfo
  {author} {\bibfnamefont {T.}~\bibnamefont {Jungwirth}}, \bibinfo {author}
  {\bibfnamefont {O.}~\bibnamefont {Gomonay}}, \bibinfo {author} {\bibfnamefont
  {J.}~\bibnamefont {Sinova}}, \bibinfo {author} {\bibfnamefont {A.~A.}\
  \bibnamefont {Sapozhnik}}, \bibinfo {author} {\bibfnamefont {H.~J.}\
  \bibnamefont {Elmers}}, \bibinfo {author} {\bibfnamefont {M.}~\bibnamefont
  {Kläui}},\ and\ \bibinfo {author} {\bibfnamefont {M.}~\bibnamefont
  {Jourdan}},\ }\bibfield  {title} {\bibinfo {title} {{Writing and reading
  antiferromagnetic Mn$_2$Au by Néel spin-orbit torques and large anisotropic
  magnetoresistance}},\ }\href {https://doi.org/10.1038/s41467-017-02780-x}
  {\bibfield  {journal} {\bibinfo  {journal} {Nature Communications}\ }\textbf
  {\bibinfo {volume} {9}},\ \bibinfo {pages} {348} (\bibinfo {year}
  {2018})}\BibitemShut {NoStop}%
\bibitem [{\citenamefont {Wadley}\ \emph {et~al.}(2018)\citenamefont {Wadley},
  \citenamefont {Reimers}, \citenamefont {Grzybowski}, \citenamefont {Andrews},
  \citenamefont {Wang}, \citenamefont {Chauhan}, \citenamefont {Gallagher},
  \citenamefont {Campion}, \citenamefont {Edmonds}, \citenamefont {Dhesi},
  \citenamefont {Maccherozzi}, \citenamefont {Novak}, \citenamefont
  {Wunderlich},\ and\ \citenamefont {Jungwirth}}]{Wadley:NNAN2018}%
  \BibitemOpen
  \bibfield  {author} {\bibinfo {author} {\bibfnamefont {P.}~\bibnamefont
  {Wadley}}, \bibinfo {author} {\bibfnamefont {S.}~\bibnamefont {Reimers}},
  \bibinfo {author} {\bibfnamefont {M.~J.}\ \bibnamefont {Grzybowski}},
  \bibinfo {author} {\bibfnamefont {C.}~\bibnamefont {Andrews}}, \bibinfo
  {author} {\bibfnamefont {M.}~\bibnamefont {Wang}}, \bibinfo {author}
  {\bibfnamefont {J.~S.}\ \bibnamefont {Chauhan}}, \bibinfo {author}
  {\bibfnamefont {B.~L.}\ \bibnamefont {Gallagher}}, \bibinfo {author}
  {\bibfnamefont {R.~P.}\ \bibnamefont {Campion}}, \bibinfo {author}
  {\bibfnamefont {K.~W.}\ \bibnamefont {Edmonds}}, \bibinfo {author}
  {\bibfnamefont {S.~S.}\ \bibnamefont {Dhesi}}, \bibinfo {author}
  {\bibfnamefont {F.}~\bibnamefont {Maccherozzi}}, \bibinfo {author}
  {\bibfnamefont {V.}~\bibnamefont {Novak}}, \bibinfo {author} {\bibfnamefont
  {J.}~\bibnamefont {Wunderlich}},\ and\ \bibinfo {author} {\bibfnamefont
  {T.}~\bibnamefont {Jungwirth}},\ }\bibfield  {title} {\bibinfo {title}
  {Current polarity-dependent manipulation of antiferromagnetic domains},\
  }\href {https://doi.org/10.1038/s41565-018-0079-1} {\bibfield  {journal}
  {\bibinfo  {journal} {Nature Nanotechnology}\ }\textbf {\bibinfo {volume}
  {13}},\ \bibinfo {pages} {362} (\bibinfo {year} {2018})}\BibitemShut
  {NoStop}%
\bibitem [{\citenamefont {Lebrun}\ \emph {et~al.}(2018)\citenamefont {Lebrun},
  \citenamefont {Ross}, \citenamefont {Bender}, \citenamefont {Qaiumzadeh},
  \citenamefont {Baldrati}, \citenamefont {Cramer}, \citenamefont {Brataas},
  \citenamefont {Duine},\ and\ \citenamefont {Kläui}}]{Lebrun:NATURE2018}%
  \BibitemOpen
  \bibfield  {author} {\bibinfo {author} {\bibfnamefont {R.}~\bibnamefont
  {Lebrun}}, \bibinfo {author} {\bibfnamefont {A.}~\bibnamefont {Ross}},
  \bibinfo {author} {\bibfnamefont {S.~A.}\ \bibnamefont {Bender}}, \bibinfo
  {author} {\bibfnamefont {A.}~\bibnamefont {Qaiumzadeh}}, \bibinfo {author}
  {\bibfnamefont {L.}~\bibnamefont {Baldrati}}, \bibinfo {author}
  {\bibfnamefont {J.}~\bibnamefont {Cramer}}, \bibinfo {author} {\bibfnamefont
  {A.}~\bibnamefont {Brataas}}, \bibinfo {author} {\bibfnamefont {R.~A.}\
  \bibnamefont {Duine}},\ and\ \bibinfo {author} {\bibfnamefont
  {M.}~\bibnamefont {Kläui}},\ }\bibfield  {title} {\bibinfo {title} {Tunable
  long-distance spin transport in a crystalline antiferromagnetic iron oxide},\
  }\href {https://doi.org/10.1038/s41586-018-0490-7} {\bibfield  {journal}
  {\bibinfo  {journal} {Nature}\ }\textbf {\bibinfo {volume} {561}},\ \bibinfo
  {pages} {222} (\bibinfo {year} {2018})}\BibitemShut {NoStop}%
\bibitem [{\citenamefont {Tserkovnyak}\ \emph {et~al.}(2002)\citenamefont
  {Tserkovnyak}, \citenamefont {Brataas},\ and\ \citenamefont
  {Bauer}}]{Tserkovnyak:PRL2002}%
  \BibitemOpen
  \bibfield  {author} {\bibinfo {author} {\bibfnamefont {Y.}~\bibnamefont
  {Tserkovnyak}}, \bibinfo {author} {\bibfnamefont {A.}~\bibnamefont
  {Brataas}},\ and\ \bibinfo {author} {\bibfnamefont {G.~E.~W.}\ \bibnamefont
  {Bauer}},\ }\bibfield  {title} {\bibinfo {title} {Enhanced gilbert damping in
  thin ferromagnetic films},\ }\href@noop {} {\bibfield  {journal} {\bibinfo
  {journal} {Physical Review Letters}\ }\textbf {\bibinfo {volume} {88}},\
  \bibinfo {pages} {117601} (\bibinfo {year} {2002})}\BibitemShut {NoStop}%
\bibitem [{\citenamefont {Brataas}\ \emph {et~al.}(2002)\citenamefont
  {Brataas}, \citenamefont {Tserkovnyak}, \citenamefont {Bauer},\ and\
  \citenamefont {Halperin}}]{Brataas:PRB2002}%
  \BibitemOpen
  \bibfield  {author} {\bibinfo {author} {\bibfnamefont {A.}~\bibnamefont
  {Brataas}}, \bibinfo {author} {\bibfnamefont {Y.}~\bibnamefont
  {Tserkovnyak}}, \bibinfo {author} {\bibfnamefont {G.~E.~W.}\ \bibnamefont
  {Bauer}},\ and\ \bibinfo {author} {\bibfnamefont {B.~I.}\ \bibnamefont
  {Halperin}},\ }\bibfield  {title} {\bibinfo {title} {Spin battery operated by
  ferromagnetic resonance},\ }\href
  {http://link.aps.org/doi/10.1103/PhysRevB.66.060404} {\bibfield  {journal}
  {\bibinfo  {journal} {Physical Review B}\ }\textbf {\bibinfo {volume} {66}},\
  \bibinfo {pages} {060404} (\bibinfo {year} {2002})}\BibitemShut {NoStop}%
\bibitem [{\citenamefont {Tserkovnyak}\ \emph {et~al.}(2005)\citenamefont
  {Tserkovnyak}, \citenamefont {Brataas}, \citenamefont {Bauer},\ and\
  \citenamefont {Halperin}}]{Tserkovnyak:RMP2005}%
  \BibitemOpen
  \bibfield  {author} {\bibinfo {author} {\bibfnamefont {Y.}~\bibnamefont
  {Tserkovnyak}}, \bibinfo {author} {\bibfnamefont {A.}~\bibnamefont
  {Brataas}}, \bibinfo {author} {\bibfnamefont {G.~E.~W.}\ \bibnamefont
  {Bauer}},\ and\ \bibinfo {author} {\bibfnamefont {B.~I.}\ \bibnamefont
  {Halperin}},\ }\bibfield  {title} {\bibinfo {title} {Nonlocal magnetization
  dynamics in ferromagnetic heterostructures},\ }\href
  {http://link.aps.org/abstract/RMP/v77/p1375} {\bibfield  {journal} {\bibinfo
  {journal} {Reviews of Modern Physics}\ }\textbf {\bibinfo {volume} {77}},\
  \bibinfo {pages} {1375} (\bibinfo {year} {2005})}\BibitemShut {NoStop}%
\bibitem [{\citenamefont {Cheng}\ \emph {et~al.}(2014)\citenamefont {Cheng},
  \citenamefont {Xiao}, \citenamefont {Niu},\ and\ \citenamefont
  {Brataas}}]{Cheng:PRL2014}%
  \BibitemOpen
  \bibfield  {author} {\bibinfo {author} {\bibfnamefont {R.}~\bibnamefont
  {Cheng}}, \bibinfo {author} {\bibfnamefont {J.}~\bibnamefont {Xiao}},
  \bibinfo {author} {\bibfnamefont {Q.}~\bibnamefont {Niu}},\ and\ \bibinfo
  {author} {\bibfnamefont {A.}~\bibnamefont {Brataas}},\ }\bibfield  {title}
  {\bibinfo {title} {{Spin Pumping and Spin-Transfer Torques in
  Antiferromagnets}},\ }\href {<Go to ISI>://WOS:000342692200015} {\bibfield
  {journal} {\bibinfo  {journal} {Physical Review Letters}\ }\textbf {\bibinfo
  {volume} {113}} (\bibinfo {year} {2014})}\BibitemShut {NoStop}%
\bibitem [{\citenamefont {Li}\ \emph {et~al.}(2020)\citenamefont {Li},
  \citenamefont {Wilson}, \citenamefont {Cheng}, \citenamefont {Lohmann},
  \citenamefont {Kavand}, \citenamefont {Yuan}, \citenamefont {Aldosary},
  \citenamefont {Agladze}, \citenamefont {Wei}, \citenamefont {Sherwin},\ and\
  \citenamefont {Shi}}]{Li:Nature2020}%
  \BibitemOpen
  \bibfield  {author} {\bibinfo {author} {\bibfnamefont {J.}~\bibnamefont
  {Li}}, \bibinfo {author} {\bibfnamefont {C.~B.}\ \bibnamefont {Wilson}},
  \bibinfo {author} {\bibfnamefont {R.}~\bibnamefont {Cheng}}, \bibinfo
  {author} {\bibfnamefont {M.}~\bibnamefont {Lohmann}}, \bibinfo {author}
  {\bibfnamefont {M.}~\bibnamefont {Kavand}}, \bibinfo {author} {\bibfnamefont
  {W.}~\bibnamefont {Yuan}}, \bibinfo {author} {\bibfnamefont {M.}~\bibnamefont
  {Aldosary}}, \bibinfo {author} {\bibfnamefont {N.}~\bibnamefont {Agladze}},
  \bibinfo {author} {\bibfnamefont {P.}~\bibnamefont {Wei}}, \bibinfo {author}
  {\bibfnamefont {M.~S.}\ \bibnamefont {Sherwin}},\ and\ \bibinfo {author}
  {\bibfnamefont {J.}~\bibnamefont {Shi}},\ }\bibfield  {title} {\bibinfo
  {title} {Spin current from sub-terahertz-generated antiferromagnetic
  magnons},\ }\href {https://doi.org/10.1038/s41586-020-1950-4} {\bibfield
  {journal} {\bibinfo  {journal} {Nature}\ }\textbf {\bibinfo {volume} {578}},\
  \bibinfo {pages} {70} (\bibinfo {year} {2020})}\BibitemShut {NoStop}%
\bibitem [{\citenamefont {Vaidya}\ \emph {et~al.}(2020)\citenamefont {Vaidya},
  \citenamefont {Morley}, \citenamefont {van Tol}, \citenamefont {Liu},
  \citenamefont {Cheng}, \citenamefont {Brataas}, \citenamefont {Lederman},\
  and\ \citenamefont {del Barco}}]{Vaidya:SCIENCE2020}%
  \BibitemOpen
  \bibfield  {author} {\bibinfo {author} {\bibfnamefont {P.}~\bibnamefont
  {Vaidya}}, \bibinfo {author} {\bibfnamefont {S.~A.}\ \bibnamefont {Morley}},
  \bibinfo {author} {\bibfnamefont {J.}~\bibnamefont {van Tol}}, \bibinfo
  {author} {\bibfnamefont {Y.}~\bibnamefont {Liu}}, \bibinfo {author}
  {\bibfnamefont {R.}~\bibnamefont {Cheng}}, \bibinfo {author} {\bibfnamefont
  {A.}~\bibnamefont {Brataas}}, \bibinfo {author} {\bibfnamefont
  {D.}~\bibnamefont {Lederman}},\ and\ \bibinfo {author} {\bibfnamefont
  {E.}~\bibnamefont {del Barco}},\ }\bibfield  {title} {\bibinfo {title}
  {Subterahertz spin pumping from an insulating antiferromagnet},\ }\href
  {https://doi.org/10.1126/science.aaz4247} {\bibfield  {journal} {\bibinfo
  {journal} {Science}\ }\textbf {\bibinfo {volume} {368}},\ \bibinfo {pages}
  {160} (\bibinfo {year} {2020})}\BibitemShut {NoStop}%
\bibitem [{\citenamefont {Ohno}\ \emph {et~al.}(2000)\citenamefont {Ohno},
  \citenamefont {Chiba}, \citenamefont {Matsukura}, \citenamefont {Omiya},
  \citenamefont {Abe}, \citenamefont {Dietl}, \citenamefont {Ohno},\ and\
  \citenamefont {Ohtani}}]{Ohno:NATURE2000}%
  \BibitemOpen
  \bibfield  {author} {\bibinfo {author} {\bibfnamefont {H.}~\bibnamefont
  {Ohno}}, \bibinfo {author} {\bibfnamefont {D.}~\bibnamefont {Chiba}},
  \bibinfo {author} {\bibfnamefont {F.}~\bibnamefont {Matsukura}}, \bibinfo
  {author} {\bibfnamefont {T.}~\bibnamefont {Omiya}}, \bibinfo {author}
  {\bibfnamefont {E.}~\bibnamefont {Abe}}, \bibinfo {author} {\bibfnamefont
  {T.}~\bibnamefont {Dietl}}, \bibinfo {author} {\bibfnamefont
  {Y.}~\bibnamefont {Ohno}},\ and\ \bibinfo {author} {\bibfnamefont
  {K.}~\bibnamefont {Ohtani}},\ }\bibfield  {title} {\bibinfo {title}
  {Electric-field control of ferromagnetism},\ }\href
  {http://dx.doi.org/10.1038/35050040} {\bibfield  {journal} {\bibinfo
  {journal} {Nature}\ }\textbf {\bibinfo {volume} {408}},\ \bibinfo {pages}
  {944} (\bibinfo {year} {2000})}\BibitemShut {NoStop}%
\bibitem [{\citenamefont {Jungwirth}\ \emph {et~al.}(2006)\citenamefont
  {Jungwirth}, \citenamefont {Sinova}, \citenamefont {Mašek}, \citenamefont
  {Kučera},\ and\ \citenamefont {MacDonald}}]{Jungwirth:RMP2006}%
  \BibitemOpen
  \bibfield  {author} {\bibinfo {author} {\bibfnamefont {T.}~\bibnamefont
  {Jungwirth}}, \bibinfo {author} {\bibfnamefont {J.}~\bibnamefont {Sinova}},
  \bibinfo {author} {\bibfnamefont {J.}~\bibnamefont {Mašek}}, \bibinfo
  {author} {\bibfnamefont {J.}~\bibnamefont {Kučera}},\ and\ \bibinfo {author}
  {\bibfnamefont {A.~H.}\ \bibnamefont {MacDonald}},\ }\bibfield  {title}
  {\bibinfo {title} {{Theory of ferromagnetic (III,Mn)V semiconductors}},\
  }\href {http://link.aps.org/doi/10.1103/RevModPhys.78.809} {\bibfield
  {journal} {\bibinfo  {journal} {Reviews of Modern Physics}\ }\textbf
  {\bibinfo {volume} {78}},\ \bibinfo {pages} {809} (\bibinfo {year}
  {2006})}\BibitemShut {NoStop}%
\bibitem [{\citenamefont {Tanaka}(2020)}]{Tanaka:JJAP2020}%
  \BibitemOpen
  \bibfield  {author} {\bibinfo {author} {\bibfnamefont {M.}~\bibnamefont
  {Tanaka}},\ }\bibfield  {title} {\bibinfo {title} {Recent progress in
  ferromagnetic semiconductors and spintronics devices},\ }\href
  {https://doi.org/10.35848/1347-4065/abcadc} {\bibfield  {journal} {\bibinfo
  {journal} {Japanese Journal of Applied Physics}\ }\textbf {\bibinfo {volume}
  {60}},\ \bibinfo {pages} {010101} (\bibinfo {year} {2020})}\BibitemShut
  {NoStop}%
\bibitem [{\citenamefont {Teranishi}(1961)}]{teranishi1961}%
  \BibitemOpen
  \bibfield  {author} {\bibinfo {author} {\bibfnamefont {T.}~\bibnamefont
  {Teranishi}},\ }\bibfield  {title} {\bibinfo {title} {Magnetic and electric
  properties of chalcopyrite},\ }\href@noop {} {\bibfield  {journal} {\bibinfo
  {journal} {J. Phys. Soc. Japan}\ }\textbf {\bibinfo {volume} {16}},\ \bibinfo
  {pages} {1881} (\bibinfo {year} {1961})}\BibitemShut {NoStop}%
\bibitem [{\citenamefont {{\v{S}}mejkal}\ \emph {et~al.}(2020)\citenamefont
  {{\v{S}}mejkal}, \citenamefont {Gonz{\'a}lez-Hern{\'a}ndez}, \citenamefont
  {Jungwirth},\ and\ \citenamefont {Sinova}}]{vsmejkal2020crystal}%
  \BibitemOpen
  \bibfield  {author} {\bibinfo {author} {\bibfnamefont {L.}~\bibnamefont
  {{\v{S}}mejkal}}, \bibinfo {author} {\bibfnamefont {R.}~\bibnamefont
  {Gonz{\'a}lez-Hern{\'a}ndez}}, \bibinfo {author} {\bibfnamefont
  {T.}~\bibnamefont {Jungwirth}},\ and\ \bibinfo {author} {\bibfnamefont
  {J.}~\bibnamefont {Sinova}},\ }\bibfield  {title} {\bibinfo {title} {{Crystal
  time-reversal symmetry breaking and spontaneous Hall effect in collinear
  antiferromagnets}},\ }\href {https://doi.org/10.1126/sciadv.aaz8809}
  {\bibfield  {journal} {\bibinfo  {journal} {Science Advances}\ }\textbf
  {\bibinfo {volume} {6}},\ \bibinfo {pages} {eaaz8809} (\bibinfo {year}
  {2020})}\BibitemShut {NoStop}%
\bibitem [{\citenamefont {Yuan}\ \emph {et~al.}(2020)\citenamefont {Yuan},
  \citenamefont {Wang}, \citenamefont {Luo}, \citenamefont {Rashba},\ and\
  \citenamefont {Zunger}}]{yuan2020giant}%
  \BibitemOpen
  \bibfield  {author} {\bibinfo {author} {\bibfnamefont {L.-D.}\ \bibnamefont
  {Yuan}}, \bibinfo {author} {\bibfnamefont {Z.}~\bibnamefont {Wang}}, \bibinfo
  {author} {\bibfnamefont {J.-W.}\ \bibnamefont {Luo}}, \bibinfo {author}
  {\bibfnamefont {E.~I.}\ \bibnamefont {Rashba}},\ and\ \bibinfo {author}
  {\bibfnamefont {A.}~\bibnamefont {Zunger}},\ }\bibfield  {title} {\bibinfo
  {title} {{Giant momentum-dependent spin splitting in centrosymmetric low-Z
  antiferromagnets}},\ }\href@noop {} {\bibfield  {journal} {\bibinfo
  {journal} {Physical Review B}\ }\textbf {\bibinfo {volume} {102}},\ \bibinfo
  {pages} {014422} (\bibinfo {year} {2020})}\BibitemShut {NoStop}%
\bibitem [{\citenamefont {Kittel}\ \emph {et~al.}(1996)\citenamefont {Kittel},
  \citenamefont {McEuen},\ and\ \citenamefont
  {McEuen}}]{kittel1996introduction}%
  \BibitemOpen
  \bibfield  {author} {\bibinfo {author} {\bibfnamefont {C.}~\bibnamefont
  {Kittel}}, \bibinfo {author} {\bibfnamefont {P.}~\bibnamefont {McEuen}},\
  and\ \bibinfo {author} {\bibfnamefont {P.}~\bibnamefont {McEuen}},\
  }\href@noop {} {\emph {\bibinfo {title} {Introduction to Solid State
  Physics}}},\ Vol.~\bibinfo {volume} {8}\ (\bibinfo  {publisher} {Wiley New
  York},\ \bibinfo {year} {1996})\BibitemShut {NoStop}%
\bibitem [{\citenamefont {Łażewski}\ \emph {et~al.}(2004)\citenamefont
  {Łażewski}, \citenamefont {Neumann},\ and\ \citenamefont
  {Parlinski}}]{Lazewski:PRB2004}%
  \BibitemOpen
  \bibfield  {author} {\bibinfo {author} {\bibfnamefont {J.}~\bibnamefont
  {Łażewski}}, \bibinfo {author} {\bibfnamefont {H.}~\bibnamefont
  {Neumann}},\ and\ \bibinfo {author} {\bibfnamefont {K.}~\bibnamefont
  {Parlinski}},\ }\bibfield  {title} {\bibinfo {title} {{Ab initio
  characterization of magnetic ${\mathrm{CuFeS}}_{2}$}},\ }\href
  {https://doi.org/10.1103/PhysRevB.70.195206} {\bibfield  {journal} {\bibinfo
  {journal} {Physical Review B}\ }\textbf {\bibinfo {volume} {70}},\ \bibinfo
  {pages} {195206} (\bibinfo {year} {2004})}\BibitemShut {NoStop}%
\bibitem [{\citenamefont {Park}\ \emph {et~al.}(2019)\citenamefont {Park},
  \citenamefont {Xia},\ and\ \citenamefont
  {Ozoli{\c{n}}{\v{s}}}}]{Park:JAP2019}%
  \BibitemOpen
  \bibfield  {author} {\bibinfo {author} {\bibfnamefont {J.}~\bibnamefont
  {Park}}, \bibinfo {author} {\bibfnamefont {Y.}~\bibnamefont {Xia}},\ and\
  \bibinfo {author} {\bibfnamefont {V.}~\bibnamefont {Ozoli{\c{n}}{\v{s}}}},\
  }\bibfield  {title} {\bibinfo {title} {{First-principles assessment of
  thermoelectric properties of CuFeS$_2$}},\ }\href
  {https://doi.org/https://doi.org/10.1063/1.5088165} {\bibfield  {journal}
  {\bibinfo  {journal} {Journal of Applied Physics}\ }\textbf {\bibinfo
  {volume} {125}},\ \bibinfo {pages} {125102} (\bibinfo {year}
  {2019})}\BibitemShut {NoStop}%
\bibitem [{\citenamefont {Takaki}\ \emph {et~al.}(2019)\citenamefont {Takaki},
  \citenamefont {Kobayashi}, \citenamefont {Shimono}, \citenamefont
  {Kobayashi}, \citenamefont {Hirose}, \citenamefont {Tsujii},\ and\
  \citenamefont {Mori}}]{Hirokazu:JJAP2019}%
  \BibitemOpen
  \bibfield  {author} {\bibinfo {author} {\bibfnamefont {H.}~\bibnamefont
  {Takaki}}, \bibinfo {author} {\bibfnamefont {K.}~\bibnamefont {Kobayashi}},
  \bibinfo {author} {\bibfnamefont {M.}~\bibnamefont {Shimono}}, \bibinfo
  {author} {\bibfnamefont {N.}~\bibnamefont {Kobayashi}}, \bibinfo {author}
  {\bibfnamefont {K.}~\bibnamefont {Hirose}}, \bibinfo {author} {\bibfnamefont
  {N.}~\bibnamefont {Tsujii}},\ and\ \bibinfo {author} {\bibfnamefont
  {T.}~\bibnamefont {Mori}},\ }\bibfield  {title} {\bibinfo {title} {{Seebeck
  coefficients in CuFeS$_2$ thin films by first-principles calculations}},\
  }\href {https://doi.org/10.7567/1347-4065/ab147c} {\bibfield  {journal}
  {\bibinfo  {journal} {Japanese Journal of Applied Physics}\ }\textbf
  {\bibinfo {volume} {58}},\ \bibinfo {pages} {SIIB01} (\bibinfo {year}
  {2019})}\BibitemShut {NoStop}%
\bibitem [{\citenamefont {Takaki}\ \emph {et~al.}(2017)\citenamefont {Takaki},
  \citenamefont {Kobayashi}, \citenamefont {Shimono}, \citenamefont
  {Kobayashi}, \citenamefont {Hirose}, \citenamefont {Tsujii},\ and\
  \citenamefont {Mori}}]{Hirokazu:APL2017}%
  \BibitemOpen
  \bibfield  {author} {\bibinfo {author} {\bibfnamefont {H.}~\bibnamefont
  {Takaki}}, \bibinfo {author} {\bibfnamefont {K.}~\bibnamefont {Kobayashi}},
  \bibinfo {author} {\bibfnamefont {M.}~\bibnamefont {Shimono}}, \bibinfo
  {author} {\bibfnamefont {N.}~\bibnamefont {Kobayashi}}, \bibinfo {author}
  {\bibfnamefont {K.}~\bibnamefont {Hirose}}, \bibinfo {author} {\bibfnamefont
  {N.}~\bibnamefont {Tsujii}},\ and\ \bibinfo {author} {\bibfnamefont
  {T.}~\bibnamefont {Mori}},\ }\bibfield  {title} {\bibinfo {title}
  {{First-principles calculations of Seebeck coefficients in a magnetic
  semiconductor CuFeS$_2$}},\ }\href {https://doi.org/10.1063/1.4976574}
  {\bibfield  {journal} {\bibinfo  {journal} {Applied Physics Letters}\
  }\textbf {\bibinfo {volume} {110}},\ \bibinfo {pages} {072107} (\bibinfo
  {year} {2017})}\BibitemShut {NoStop}%
\bibitem [{\citenamefont {Zhou}\ \emph {et~al.}(2015)\citenamefont {Zhou},
  \citenamefont {Gao}, \citenamefont {Cheng}, \citenamefont {Chen},\ and\
  \citenamefont {Cai}}]{Meng:APA2015}%
  \BibitemOpen
  \bibfield  {author} {\bibinfo {author} {\bibfnamefont {M.}~\bibnamefont
  {Zhou}}, \bibinfo {author} {\bibfnamefont {X.}~\bibnamefont {Gao}}, \bibinfo
  {author} {\bibfnamefont {Y.}~\bibnamefont {Cheng}}, \bibinfo {author}
  {\bibfnamefont {X.}~\bibnamefont {Chen}},\ and\ \bibinfo {author}
  {\bibfnamefont {L.}~\bibnamefont {Cai}},\ }\bibfield  {title} {\bibinfo
  {title} {{Structural, electronic, and elastic properties of CuFeS$_2$:
  first-principles study}},\ }\href {https://doi.org/10.1007/s00339-014-8930-1}
  {\bibfield  {journal} {\bibinfo  {journal} {Applied Physics A}\ }\textbf
  {\bibinfo {volume} {118}},\ \bibinfo {pages} {1145} (\bibinfo {year}
  {2015})}\BibitemShut {NoStop}%
\bibitem [{\citenamefont {Conejeros}\ \emph {et~al.}(2015)\citenamefont
  {Conejeros}, \citenamefont {Alemany}, \citenamefont {Llunell}, \citenamefont
  {Moreira}, \citenamefont {Sánchez},\ and\ \citenamefont
  {Llanos}}]{Conejeros:IC2015}%
  \BibitemOpen
  \bibfield  {author} {\bibinfo {author} {\bibfnamefont {S.}~\bibnamefont
  {Conejeros}}, \bibinfo {author} {\bibfnamefont {P.}~\bibnamefont {Alemany}},
  \bibinfo {author} {\bibfnamefont {M.}~\bibnamefont {Llunell}}, \bibinfo
  {author} {\bibfnamefont {I.~d. P.~R.}\ \bibnamefont {Moreira}}, \bibinfo
  {author} {\bibfnamefont {V.}~\bibnamefont {Sánchez}},\ and\ \bibinfo
  {author} {\bibfnamefont {J.}~\bibnamefont {Llanos}},\ }\bibfield  {title}
  {\bibinfo {title} {{Electronic Structure and Magnetic Properties of
  CuFeS$_2$}},\ }\href {https://doi.org/10.1021/acs.inorgchem.5b00399}
  {\bibfield  {journal} {\bibinfo  {journal} {Inorganic Chemistry}\ }\textbf
  {\bibinfo {volume} {54}},\ \bibinfo {pages} {4840} (\bibinfo {year}
  {2015})}\BibitemShut {NoStop}%
\bibitem [{\citenamefont {Khaledialidusti}\ \emph {et~al.}(2019)\citenamefont
  {Khaledialidusti}, \citenamefont {Mishra},\ and\ \citenamefont
  {Barnoush}}]{khaledialidusti2019temperature}%
  \BibitemOpen
  \bibfield  {author} {\bibinfo {author} {\bibfnamefont {R.}~\bibnamefont
  {Khaledialidusti}}, \bibinfo {author} {\bibfnamefont {A.~K.}\ \bibnamefont
  {Mishra}},\ and\ \bibinfo {author} {\bibfnamefont {A.}~\bibnamefont
  {Barnoush}},\ }\bibfield  {title} {\bibinfo {title} {{Temperature-dependent
  properties of magnetic CuFeS$_2$ from first-principles calculations:
  Structure, mechanics, and thermodynamics}},\ }\href@noop {} {\bibfield
  {journal} {\bibinfo  {journal} {AIP Advances}\ }\textbf {\bibinfo {volume}
  {9}},\ \bibinfo {pages} {065021} (\bibinfo {year} {2019})}\BibitemShut
  {NoStop}%
\bibitem [{\citenamefont {Hamajima}\ \emph {et~al.}(1981)\citenamefont
  {Hamajima}, \citenamefont {Kambara}, \citenamefont {Gondaira},\ and\
  \citenamefont {Oguchi}}]{hamajima1981self}%
  \BibitemOpen
  \bibfield  {author} {\bibinfo {author} {\bibfnamefont {T.}~\bibnamefont
  {Hamajima}}, \bibinfo {author} {\bibfnamefont {T.}~\bibnamefont {Kambara}},
  \bibinfo {author} {\bibfnamefont {K.~I.}\ \bibnamefont {Gondaira}},\ and\
  \bibinfo {author} {\bibfnamefont {T.}~\bibnamefont {Oguchi}},\ }\bibfield
  {title} {\bibinfo {title} {{Self-consistent electronic structures of magnetic
  semiconductors by a discrete variational X$\alpha$ calculation. III.
  Chalcopyrite CuFeS$_2$}},\ }\href@noop {} {\bibfield  {journal} {\bibinfo
  {journal} {Physical Review B}\ }\textbf {\bibinfo {volume} {24}},\ \bibinfo
  {pages} {3349} (\bibinfo {year} {1981})}\BibitemShut {NoStop}%
\bibitem [{\citenamefont {Edelbro}\ \emph {et~al.}(2003)\citenamefont
  {Edelbro}, \citenamefont {Sandstr{\"o}m},\ and\ \citenamefont
  {Paul}}]{edelbro2003full}%
  \BibitemOpen
  \bibfield  {author} {\bibinfo {author} {\bibfnamefont {R.}~\bibnamefont
  {Edelbro}}, \bibinfo {author} {\bibfnamefont {{\AA}.}~\bibnamefont
  {Sandstr{\"o}m}},\ and\ \bibinfo {author} {\bibfnamefont {J.}~\bibnamefont
  {Paul}},\ }\bibfield  {title} {\bibinfo {title} {Full potential calculations
  on the electron bandstructures of sphalerite, pyrite and chalcopyrite},\
  }\href@noop {} {\bibfield  {journal} {\bibinfo  {journal} {Applied Surface
  Science}\ }\textbf {\bibinfo {volume} {206}},\ \bibinfo {pages} {300}
  (\bibinfo {year} {2003})}\BibitemShut {NoStop}%
\bibitem [{\citenamefont {de~Oliveira}\ and\ \citenamefont
  {Duarte}(2010)}]{de2010disulphide}%
  \BibitemOpen
  \bibfield  {author} {\bibinfo {author} {\bibfnamefont {C.}~\bibnamefont
  {de~Oliveira}}\ and\ \bibinfo {author} {\bibfnamefont {H.~A.}\ \bibnamefont
  {Duarte}},\ }\bibfield  {title} {\bibinfo {title} {{Disulphide and metal
  sulphide formation on the reconstructed (001) surface of chalcopyrite: A DFT
  study}},\ }\href@noop {} {\bibfield  {journal} {\bibinfo  {journal} {Applied
  Surface Science}\ }\textbf {\bibinfo {volume} {257}},\ \bibinfo {pages}
  {1319} (\bibinfo {year} {2010})}\BibitemShut {NoStop}%
\bibitem [{\citenamefont {Pauling}\ and\ \citenamefont
  {Brockway}(1932)}]{pauling1932crystal}%
  \BibitemOpen
  \bibfield  {author} {\bibinfo {author} {\bibfnamefont {L.}~\bibnamefont
  {Pauling}}\ and\ \bibinfo {author} {\bibfnamefont {L.}~\bibnamefont
  {Brockway}},\ }\bibfield  {title} {\bibinfo {title} {{The crystal structure
  of chalcopyrite CuFeS$_2$}},\ }\href@noop {} {\bibfield  {journal} {\bibinfo
  {journal} {Zeitschrift f{\"u}r Kristallographie-Crystalline Materials}\
  }\textbf {\bibinfo {volume} {82}},\ \bibinfo {pages} {188} (\bibinfo {year}
  {1932})}\BibitemShut {NoStop}%
\bibitem [{\citenamefont {Donnay}\ \emph {et~al.}(1958)\citenamefont {Donnay},
  \citenamefont {Corliss}, \citenamefont {Donnay}, \citenamefont {Elliott},\
  and\ \citenamefont {Hastings}}]{donnay1958symmetry}%
  \BibitemOpen
  \bibfield  {author} {\bibinfo {author} {\bibfnamefont {G.}~\bibnamefont
  {Donnay}}, \bibinfo {author} {\bibfnamefont {L.}~\bibnamefont {Corliss}},
  \bibinfo {author} {\bibfnamefont {J.}~\bibnamefont {Donnay}}, \bibinfo
  {author} {\bibfnamefont {N.}~\bibnamefont {Elliott}},\ and\ \bibinfo {author}
  {\bibfnamefont {J.}~\bibnamefont {Hastings}},\ }\bibfield  {title} {\bibinfo
  {title} {Symmetry of magnetic structures: magnetic structure of
  chalcopyrite},\ }\href@noop {} {\bibfield  {journal} {\bibinfo  {journal}
  {Physical Review}\ }\textbf {\bibinfo {volume} {112}},\ \bibinfo {pages}
  {1917} (\bibinfo {year} {1958})}\BibitemShut {NoStop}%
\bibitem [{\citenamefont {Bogdanov}\ and\ \citenamefont
  {Yablonski}(1989)}]{Bogdanov:ZETF1989}%
  \BibitemOpen
  \bibfield  {author} {\bibinfo {author} {\bibfnamefont {A.~N.}\ \bibnamefont
  {Bogdanov}}\ and\ \bibinfo {author} {\bibfnamefont {D.~A.}\ \bibnamefont
  {Yablonski}},\ }\bibfield  {title} {\bibinfo {title} {{Contribution to the
  theory of inhomogeneous states of magnets in the region of
  magnetic-field-induced phase transitions. Mixed state of antiferromagnets}},\
  }\href@noop {} {\bibfield  {journal} {\bibinfo  {journal} {Zh. Eksp. Teor.
  Fiz.}\ }\textbf {\bibinfo {volume} {96}},\ \bibinfo {pages} {253} (\bibinfo
  {year} {1989})}\BibitemShut {NoStop}%
\bibitem [{\citenamefont {Birss}(1966)}]{birss1966symmetry}%
  \BibitemOpen
  \bibfield  {author} {\bibinfo {author} {\bibfnamefont {R.~R.}\ \bibnamefont
  {Birss}},\ }\href@noop {} {\emph {\bibinfo {title} {Symmetry and
  magnetism}}},\ Vol.~\bibinfo {volume} {3}\ (\bibinfo  {publisher} {Elsevier
  Science \& Technology},\ \bibinfo {year} {1966})\BibitemShut {NoStop}%
\bibitem [{\citenamefont {Blöchl}(1994)}]{Blochl:PRB1994}%
  \BibitemOpen
  \bibfield  {author} {\bibinfo {author} {\bibfnamefont {P.~E.}\ \bibnamefont
  {Blöchl}},\ }\bibfield  {title} {\bibinfo {title} {Projector augmented-wave
  method},\ }\href@noop {} {\bibfield  {journal} {\bibinfo  {journal} {Physical
  Review B}\ }\textbf {\bibinfo {volume} {50}},\ \bibinfo {pages}
  {17953–17979} (\bibinfo {year} {1994})}\BibitemShut {NoStop}%
\bibitem [{\citenamefont {Kresse}\ and\ \citenamefont
  {Joubert}(1999)}]{KresseJoubert:PRB1999}%
  \BibitemOpen
  \bibfield  {author} {\bibinfo {author} {\bibfnamefont {G.}~\bibnamefont
  {Kresse}}\ and\ \bibinfo {author} {\bibfnamefont {D.}~\bibnamefont
  {Joubert}},\ }\bibfield  {title} {\bibinfo {title} {From ultrasoft
  pseudopotentials to the projector augmented-wave method},\ }\href@noop {}
  {\bibfield  {journal} {\bibinfo  {journal} {Physical Review B}\ }\textbf
  {\bibinfo {volume} {59}},\ \bibinfo {pages} {1758–1775} (\bibinfo {year}
  {1999})}\BibitemShut {NoStop}%
\bibitem [{\citenamefont {Monkhorst}\ and\ \citenamefont
  {Pack}(1976)}]{MonkhorstPack:PRB1976}%
  \BibitemOpen
  \bibfield  {author} {\bibinfo {author} {\bibfnamefont {H.~J.}\ \bibnamefont
  {Monkhorst}}\ and\ \bibinfo {author} {\bibfnamefont {J.~D.}\ \bibnamefont
  {Pack}},\ }\bibfield  {title} {\bibinfo {title} {{Special points for
  Brillouin-zone integrations}},\ }\href@noop {} {\bibfield  {journal}
  {\bibinfo  {journal} {Physical Review B}\ }\textbf {\bibinfo {volume} {13}},\
  \bibinfo {pages} {5188} (\bibinfo {year} {1976})}\BibitemShut {NoStop}%
\bibitem [{\citenamefont {Elcoro}\ \emph {et~al.}(2017)\citenamefont {Elcoro},
  \citenamefont {Bradlyn}, \citenamefont {Wang}, \citenamefont {Vergniory},
  \citenamefont {Cano}, \citenamefont {Felser}, \citenamefont {Bernevig},
  \citenamefont {Orobengoa}, \citenamefont {Flor},\ and\ \citenamefont
  {Aroyo}}]{elcoro2017double}%
  \BibitemOpen
  \bibfield  {author} {\bibinfo {author} {\bibfnamefont {L.}~\bibnamefont
  {Elcoro}}, \bibinfo {author} {\bibfnamefont {B.}~\bibnamefont {Bradlyn}},
  \bibinfo {author} {\bibfnamefont {Z.}~\bibnamefont {Wang}}, \bibinfo {author}
  {\bibfnamefont {M.~G.}\ \bibnamefont {Vergniory}}, \bibinfo {author}
  {\bibfnamefont {J.}~\bibnamefont {Cano}}, \bibinfo {author} {\bibfnamefont
  {C.}~\bibnamefont {Felser}}, \bibinfo {author} {\bibfnamefont {B.~A.}\
  \bibnamefont {Bernevig}}, \bibinfo {author} {\bibfnamefont {D.}~\bibnamefont
  {Orobengoa}}, \bibinfo {author} {\bibfnamefont {G.}~\bibnamefont {Flor}},\
  and\ \bibinfo {author} {\bibfnamefont {M.~I.}\ \bibnamefont {Aroyo}},\
  }\bibfield  {title} {\bibinfo {title} {{Double crystallographic groups and
  their representations on the Bilbao Crystallographic Server}},\ }\href@noop
  {} {\bibfield  {journal} {\bibinfo  {journal} {Journal of Applied
  Crystallography}\ }\textbf {\bibinfo {volume} {50}},\ \bibinfo {pages} {1457}
  (\bibinfo {year} {2017})}\BibitemShut {NoStop}%
\bibitem [{\citenamefont {Dudarev}\ \emph {et~al.}(1998)\citenamefont
  {Dudarev}, \citenamefont {Botton}, \citenamefont {Savrasov}, \citenamefont
  {Humphreys},\ and\ \citenamefont {Sutton}}]{Dudarev:PRB1998}%
  \BibitemOpen
  \bibfield  {author} {\bibinfo {author} {\bibfnamefont {S.~L.}\ \bibnamefont
  {Dudarev}}, \bibinfo {author} {\bibfnamefont {G.~A.}\ \bibnamefont {Botton}},
  \bibinfo {author} {\bibfnamefont {S.~Y.}\ \bibnamefont {Savrasov}}, \bibinfo
  {author} {\bibfnamefont {C.~J.}\ \bibnamefont {Humphreys}},\ and\ \bibinfo
  {author} {\bibfnamefont {A.~P.}\ \bibnamefont {Sutton}},\ }\bibfield  {title}
  {\bibinfo {title} {{Electron-energy-loss spectra and the structural stability
  of nickel oxide: An LSDA+U study}},\ }\href
  {https://doi.org/10.1103/PhysRevB.57.1505} {\bibfield  {journal} {\bibinfo
  {journal} {Physical Review B}\ }\textbf {\bibinfo {volume} {57}},\ \bibinfo
  {pages} {1505} (\bibinfo {year} {1998})}\BibitemShut {NoStop}%
\bibitem [{\citenamefont {Perdew}\ \emph {et~al.}(2008)\citenamefont {Perdew},
  \citenamefont {Ruzsinszky}, \citenamefont {Csonka}, \citenamefont {Vydrov},
  \citenamefont {Scuseria}, \citenamefont {Constantin}, \citenamefont {Zhou},\
  and\ \citenamefont {Burke}}]{PBEsol}%
  \BibitemOpen
  \bibfield  {author} {\bibinfo {author} {\bibfnamefont {J.~P.}\ \bibnamefont
  {Perdew}}, \bibinfo {author} {\bibfnamefont {A.}~\bibnamefont {Ruzsinszky}},
  \bibinfo {author} {\bibfnamefont {G.~I.}\ \bibnamefont {Csonka}}, \bibinfo
  {author} {\bibfnamefont {O.~A.}\ \bibnamefont {Vydrov}}, \bibinfo {author}
  {\bibfnamefont {G.~E.}\ \bibnamefont {Scuseria}}, \bibinfo {author}
  {\bibfnamefont {L.~A.}\ \bibnamefont {Constantin}}, \bibinfo {author}
  {\bibfnamefont {X.}~\bibnamefont {Zhou}},\ and\ \bibinfo {author}
  {\bibfnamefont {K.}~\bibnamefont {Burke}},\ }\bibfield  {title} {\bibinfo
  {title} {{Restoring the Density-Gradient Expansion for Exchange in Solids and
  Surfaces}},\ }\href@noop {} {\bibfield  {journal} {\bibinfo  {journal} {Phys.
  Rev. Lett.}\ }\textbf {\bibinfo {volume} {100}},\ \bibinfo {pages} {136406}
  (\bibinfo {year} {2008})}\BibitemShut {NoStop}%
\bibitem [{\citenamefont {Krukau}\ \emph {et~al.}(2006)\citenamefont {Krukau},
  \citenamefont {Vydrov}, \citenamefont {Izmaylov},\ and\ \citenamefont
  {Scuseria}}]{HSE06}%
  \BibitemOpen
  \bibfield  {author} {\bibinfo {author} {\bibfnamefont {A.~V.}\ \bibnamefont
  {Krukau}}, \bibinfo {author} {\bibfnamefont {O.~A.}\ \bibnamefont {Vydrov}},
  \bibinfo {author} {\bibfnamefont {A.~F.}\ \bibnamefont {Izmaylov}},\ and\
  \bibinfo {author} {\bibfnamefont {G.~E.}\ \bibnamefont {Scuseria}},\
  }\bibfield  {title} {\bibinfo {title} {Influence of the exchange screening
  parameter on the performance of screened hybrid functionals},\ }\href@noop {}
  {\bibfield  {journal} {\bibinfo  {journal} {J. Chem. Phys.}\ }\textbf
  {\bibinfo {volume} {125}},\ \bibinfo {pages} {224106} (\bibinfo {year}
  {2006})}\BibitemShut {NoStop}%
\bibitem [{\citenamefont {Goodman}\ and\ \citenamefont
  {Douglas}(1954)}]{goodman1954}%
  \BibitemOpen
  \bibfield  {author} {\bibinfo {author} {\bibfnamefont {C.~H.~L.}\
  \bibnamefont {Goodman}}\ and\ \bibinfo {author} {\bibfnamefont {R.~W.}\
  \bibnamefont {Douglas}},\ }\bibfield  {title} {\bibinfo {title} {New
  semiconducting compounds of diamond type structure},\ }\href@noop {}
  {\bibfield  {journal} {\bibinfo  {journal} {Physica}\ }\textbf {\bibinfo
  {volume} {20}},\ \bibinfo {pages} {1107} (\bibinfo {year}
  {1954})}\BibitemShut {NoStop}%
\bibitem [{\citenamefont {Engin}\ \emph {et~al.}(2011)\citenamefont {Engin},
  \citenamefont {Powell},\ and\ \citenamefont {Hull}}]{enginPowellHull2011}%
  \BibitemOpen
  \bibfield  {author} {\bibinfo {author} {\bibfnamefont {T.~E.}\ \bibnamefont
  {Engin}}, \bibinfo {author} {\bibfnamefont {A.~V.}\ \bibnamefont {Powell}},\
  and\ \bibinfo {author} {\bibfnamefont {S.}~\bibnamefont {Hull}},\ }\bibfield
  {title} {\bibinfo {title} {{A high temperature diffraction-resistance study
  of chalcopyrite, CuFeS2}},\ }\href@noop {} {\bibfield  {journal} {\bibinfo
  {journal} {J. of Solid State Chem.}\ }\textbf {\bibinfo {volume} {184}},\
  \bibinfo {pages} {2272} (\bibinfo {year} {2011})}\BibitemShut {NoStop}%
\bibitem [{\citenamefont {Xu}\ \emph {et~al.}(2020)\citenamefont {Xu},
  \citenamefont {Elcoro}, \citenamefont {Song}, \citenamefont {Wieder},
  \citenamefont {Vergniory}, \citenamefont {Regnault}, \citenamefont {Chen},
  \citenamefont {Felser},\ and\ \citenamefont {Bernevig}}]{xu2020high}%
  \BibitemOpen
  \bibfield  {author} {\bibinfo {author} {\bibfnamefont {Y.}~\bibnamefont
  {Xu}}, \bibinfo {author} {\bibfnamefont {L.}~\bibnamefont {Elcoro}}, \bibinfo
  {author} {\bibfnamefont {Z.-D.}\ \bibnamefont {Song}}, \bibinfo {author}
  {\bibfnamefont {B.~J.}\ \bibnamefont {Wieder}}, \bibinfo {author}
  {\bibfnamefont {M.}~\bibnamefont {Vergniory}}, \bibinfo {author}
  {\bibfnamefont {N.}~\bibnamefont {Regnault}}, \bibinfo {author}
  {\bibfnamefont {Y.}~\bibnamefont {Chen}}, \bibinfo {author} {\bibfnamefont
  {C.}~\bibnamefont {Felser}},\ and\ \bibinfo {author} {\bibfnamefont {B.~A.}\
  \bibnamefont {Bernevig}},\ }\bibfield  {title} {\bibinfo {title}
  {High-throughput calculations of magnetic topological materials},\
  }\href@noop {} {\bibfield  {journal} {\bibinfo  {journal} {Nature}\ }\textbf
  {\bibinfo {volume} {586}},\ \bibinfo {pages} {702} (\bibinfo {year}
  {2020})}\BibitemShut {NoStop}%
\bibitem [{\citenamefont {Elcoro}\ \emph {et~al.}(2021)\citenamefont {Elcoro},
  \citenamefont {Wieder}, \citenamefont {Song}, \citenamefont {Xu},
  \citenamefont {Bradlyn},\ and\ \citenamefont
  {Bernevig}}]{elcoro2021magnetic}%
  \BibitemOpen
  \bibfield  {author} {\bibinfo {author} {\bibfnamefont {L.}~\bibnamefont
  {Elcoro}}, \bibinfo {author} {\bibfnamefont {B.~J.}\ \bibnamefont {Wieder}},
  \bibinfo {author} {\bibfnamefont {Z.}~\bibnamefont {Song}}, \bibinfo {author}
  {\bibfnamefont {Y.}~\bibnamefont {Xu}}, \bibinfo {author} {\bibfnamefont
  {B.}~\bibnamefont {Bradlyn}},\ and\ \bibinfo {author} {\bibfnamefont {B.~A.}\
  \bibnamefont {Bernevig}},\ }\bibfield  {title} {\bibinfo {title} {Magnetic
  topological quantum chemistry},\ }\href@noop {} {\bibfield  {journal}
  {\bibinfo  {journal} {Nature communications}\ }\textbf {\bibinfo {volume}
  {12}},\ \bibinfo {pages} {1} (\bibinfo {year} {2021})}\BibitemShut {NoStop}%
\bibitem [{\citenamefont {Voon}\ and\ \citenamefont
  {Willatzen}(2009)}]{voon2009kp}%
  \BibitemOpen
  \bibfield  {author} {\bibinfo {author} {\bibfnamefont {L.~C. L.~Y.}\
  \bibnamefont {Voon}}\ and\ \bibinfo {author} {\bibfnamefont {M.}~\bibnamefont
  {Willatzen}},\ }\href@noop {} {\emph {\bibinfo {title} {The k$\cdot$p Method:
  Electronic Properties of Semiconductors}}}\ (\bibinfo  {publisher} {Springer
  Science \& Business Media},\ \bibinfo {year} {2009})\BibitemShut {NoStop}%
\bibitem [{\citenamefont {Dresselhaus}\ \emph {et~al.}(2007)\citenamefont
  {Dresselhaus}, \citenamefont {Dresselhaus},\ and\ \citenamefont
  {Jorio}}]{dresselhaus2007group}%
  \BibitemOpen
  \bibfield  {author} {\bibinfo {author} {\bibfnamefont {M.~S.}\ \bibnamefont
  {Dresselhaus}}, \bibinfo {author} {\bibfnamefont {G.}~\bibnamefont
  {Dresselhaus}},\ and\ \bibinfo {author} {\bibfnamefont {A.}~\bibnamefont
  {Jorio}},\ }\href@noop {} {\emph {\bibinfo {title} {Group theory: Application
  to the physics of condensed matter}}}\ (\bibinfo  {publisher} {Springer
  Science \& Business Media},\ \bibinfo {year} {2007})\BibitemShut {NoStop}%
\bibitem [{\citenamefont {Mikhailovskii}\ \emph {et~al.}(1990)\citenamefont
  {Mikhailovskii}, \citenamefont {Polubotko}, \citenamefont {Prochukhan},\ and\
  \citenamefont {Rud}}]{Mikhailovskii:PSSB1990}%
  \BibitemOpen
  \bibfield  {author} {\bibinfo {author} {\bibfnamefont {A.~P.}\ \bibnamefont
  {Mikhailovskii}}, \bibinfo {author} {\bibfnamefont {A.~M.}\ \bibnamefont
  {Polubotko}}, \bibinfo {author} {\bibfnamefont {V.~D.}\ \bibnamefont
  {Prochukhan}},\ and\ \bibinfo {author} {\bibfnamefont {Y.~V.}\ \bibnamefont
  {Rud}},\ }\bibfield  {title} {\bibinfo {title} {{Gapless State in CuFeS2}},\
  }\href {https://doi.org/10.1002/pssb.2221580122} {\bibfield  {journal}
  {\bibinfo  {journal} {physica status solidi (b)}\ }\textbf {\bibinfo {volume}
  {158}},\ \bibinfo {pages} {229} (\bibinfo {year} {1990})}\BibitemShut
  {NoStop}%
\bibitem [{\citenamefont {Luttinger}\ and\ \citenamefont
  {Kohn}(1955)}]{luttinger1955motion}%
  \BibitemOpen
  \bibfield  {author} {\bibinfo {author} {\bibfnamefont {J.~M.}\ \bibnamefont
  {Luttinger}}\ and\ \bibinfo {author} {\bibfnamefont {W.}~\bibnamefont
  {Kohn}},\ }\bibfield  {title} {\bibinfo {title} {{Motion of Electrons and
  Holes in Perturbed Periodic Fields}},\ }\href@noop {} {\bibfield  {journal}
  {\bibinfo  {journal} {Physical Review}\ }\textbf {\bibinfo {volume} {97}},\
  \bibinfo {pages} {869} (\bibinfo {year} {1955})}\BibitemShut {NoStop}%
\bibitem [{\citenamefont {Luttinger}(1956)}]{luttinger1956quantum}%
  \BibitemOpen
  \bibfield  {author} {\bibinfo {author} {\bibfnamefont {J.~M.}\ \bibnamefont
  {Luttinger}},\ }\bibfield  {title} {\bibinfo {title} {{Quantum theory of
  cyclotron resonance in semiconductors: General theory}},\ }\href@noop {}
  {\bibfield  {journal} {\bibinfo  {journal} {Physical review}\ }\textbf
  {\bibinfo {volume} {102}},\ \bibinfo {pages} {1030} (\bibinfo {year}
  {1956})}\BibitemShut {NoStop}%
\bibitem [{\citenamefont {Seeger}(2013)}]{seeger2013semiconductor}%
  \BibitemOpen
  \bibfield  {author} {\bibinfo {author} {\bibfnamefont {K.}~\bibnamefont
  {Seeger}},\ }\href@noop {} {\emph {\bibinfo {title} {Semiconductor
  physics}}}\ (\bibinfo  {publisher} {Springer Science \& Business Media},\
  \bibinfo {year} {2013})\BibitemShut {NoStop}%
\end{thebibliography}%

\end{document}